\documentclass[journal]{IEEEtran}
\usepackage{packages}

\begin{document}
\title{Feedforward Nonlinear Equalizer for Short- to Medium-Reach Wireline Links}
\author{Kunmo~Kim,~\IEEEmembership{Member,~IEEE,} Paul~Kwon,~\IEEEmembership{Member,~IEEE,} \\Elad~Alon,~\IEEEmembership{Fellow,~IEEE,} and~Ali~M.~Niknejad,~\IEEEmembership{Fellow,~IEEE}
\thanks{This work was supported in part by the Qualcomm Innovation Fellowship and the ComSenTer, a research center part of the Semiconductor Research Corporation (SRC) Program Joint University Microelectronics Program (JUMP) under Grant 044077 (Corresponding author: Kunmo Kim).
Kunmo Kim and Paul Kwon contributed equally to this work.

The authors were with the Berkeley Wireless Research Center, University of California at Berkeley, CA 94704 USA (e-mail: kunmok@berkeley.edu).
}}

\markboth{IEEE Transactions on Circuits and Systems-I: Regular Papers}%
{How to Use the IEEEtran \LaTeX \ Templates}

\maketitle
\submittednotice

\begin{abstract} 
\label{sec:abstract}
This paper presents a feedforward nonlinear equalizer (FFNE) framework for short- to medium-reach wireline links that removes the feedback-timing bottleneck of decision-feedback equalizers (DFEs) while approaching the noise-margin advantage within a characterized operating region. The proposed FFNE reduces short-window maximum-likelihood sequence estimation to a compact binary decision rule, enabling a low-complexity feedforward realization without transmitter-side encoding. For the single-postcursor NRZ case, the mathematical foundation, hardware implementation, tap adaptation, statistical analysis, and equalization limit relative to an ideal 1-tap DFE are established. A window-length-3 FFNE quantifies the performance-complexity tradeoff of longer sequence windows. The framework is further extended to PAM-4 modulation and simultaneous precursor/postcursor equalization through a pattern-detection-based FFNE (PD-FFNE), which outperforms conventional FFE+DFE baselines under representative channel conditions.

\end{abstract}

\begin{IEEEkeywords}
Nonlinear Equalizer, intersymbol interference, ISI, statistical analysis, equalizer, adaptation, serial link, SerDes, wireline.
\end{IEEEkeywords}

\section{Introduction} 
\label{sec:introduction}
\IEEEPARstart{T}{he} rapid growth in Large Language Model training and distributed AI computing continues to increase bandwidth demand in datacenter serial links. In short-to-medium-reach links, moderate frequency-dependent loss can produce severe inter-symbol interference (ISI), while receiver power and implementation complexity remain tightly constrained. For moderate-loss channels, typically 20--30~dB at Nyquist, Analog-and-Mixed-Signal (AMS) receivers require additional current to close the Decision Feedback Equalizer (DFE) feedback loop, whereas high-complexity Digital Signal Processing (DSP) solutions can be difficult to justify. Thus, a nonlinear equalizer is needed that approaches DFE-like performance without the DFE timing bottleneck or the hardware cost of conventional DSP-based detectors.

Existing approaches leave an important gap in this design space. Feedforward equalization (FFE) is readily parallelized but amplifies high-frequency noise and crosstalk, whereas DFE cancels postcursor ISI without noise enhancement but requires the first feedback tap to close within one unit interval. DSP-based nonlinear equalizers, such as the decision feedforward equalizer (DFFE) \cite{dffe} and sliding-block DFE (SB-DFE) \cite{Bailey2022, kunmo_sbdfe}, relax this timing constraint but incur substantial complexity and latency.

Several lower-complexity AMS-oriented alternatives have also been proposed. Loop-unrolled DFE \cite{dfe_loopunroll, vlad_dfe, turker_dfe, ihp_dfe, dfe_loopunroll_ibm, xilinx_dfe}, look-ahead DFE with 1+D pulse shaping \cite{broadcom_dfe}, partial-response equalization \cite{ibm_100gbps}, partially unrolled DFE \cite{pudfe}, and sequence-detection approaches \cite{aurangozeb_pam4, yusang_dicode, javadi_nrz} reduce some implementation costs, but require exponential comparator scaling, channel-dependent mode switching, careful Tx/Rx co-optimization, reduced noise margin, or transmitter-side line coding. Therefore, existing approaches do not simultaneously avoid feedback-loop timing, transmitter-side encoding, and channel-dependent co-optimization constraints.

This paper presents a Feedforward Nonlinear Equalizer (FFNE) framework to address this gap. The proposed FFNE performs nonlinear ISI cancellation in a fully feedforward datapath, eliminating the DFE feedback-loop timing bottleneck while supporting compact AMS or reduced-complexity DSP implementation. Unlike encoding-based approaches, it requires no transmitter-side encoding and therefore preserves backward compatibility with conventional SerDes links.

This work frames FFNE as a performance-to-complexity design space. The Win-2 FFNE is derived from short-window MLSE and reduced to a compact binary decision rule, with practical tap adaptation and statistical BER analysis. The analysis enables comparison with an ideal 1-tap DFE and reveals a degradation boundary at $h_1 \approx 0.293h_0$. The framework extends to Win-3 FFNE, PAM-4 signaling, and pattern-detection-based FFNE (PD-FFNE) for hardware-scaling tradeoff analysis and joint precursor/postcursor ISI mitigation. PD-FFNE outperforms conventional FFE+DFE without Tx precoding.

\section{Single-tap ($h_1$) Feedforward Nonlinear ISI equalizer for NRZ Modulation} \label{sec:sectionII}

Although \cite{emami_ffne} presented a related feedforward equalizer topology, it did not address its adaptation methodology, statistical behavior, or theoretical performance limits. This section develops an implementation-oriented Win-2 FFNE formulation and clarifies its equalization limit, together with a higher-complexity extension.

\subsection{Mathematical Foundation} \label{sec2:subsec1}

\begin{figure}[!t]
    \centering
    \includegraphics[width=0.9\columnwidth]{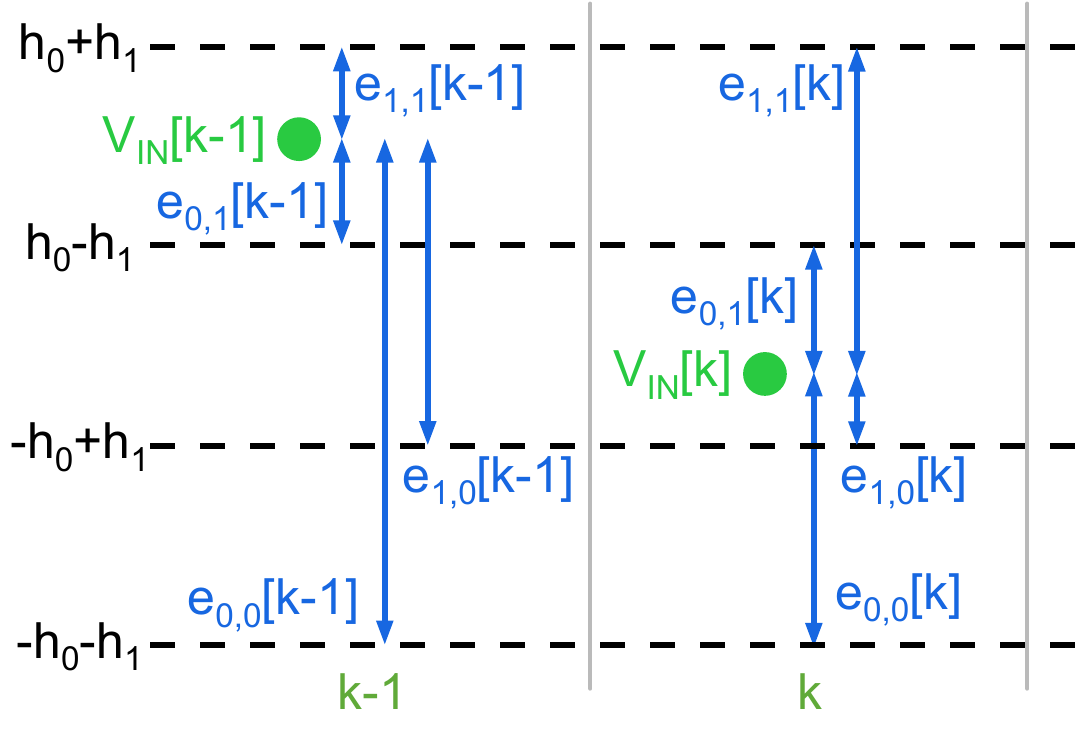}
    \caption{Ideal voltage states (black) and received voltage samples (green) for a channel with coefficients [$h_0,h_1$] in NRZ modulation.}
    \label{fig:trellis}
\end{figure}

Let $a[k]$ and $\hat{a}[k]$ represent the transmitted and detected symbols at time index $k$, respectively. For $M$-ary PAM signaling, we use the binary $a[k] \in \{0, 1, \cdots, M-1\}$ and polar $a[k] \in \{-1, \cdots, 1\}$ forms interchangeably.  The received sample $V_{\mathrm{IN}}[k]$ is defined with respect to the binary form of $a[k]$ as:
\begin{equation}
    V_{\mathrm{IN}}[k]\eqdef \sum_{i=-\infty}^\infty{h_i\left(\frac{2a[k-i]}{M-1}-1\right)}+n[k]
\end{equation}
where $h_i$ represents the ISI coefficients of the channel, and $n[k]$ is zero-mean additive white Gaussian noise (AWGN) with variance $\sigma_n^2$. The parenthesized term maps the binary form to the polar form, yielding $\{-1,+1\}$ for $M=2$ and $\{-1,-\tfrac{1}{3},+\tfrac{1}{3},+1\}$ for $M=4$. For the remainder of this section, we adopt two working assumptions: NRZ signaling ($M=2$) and a channel dominated by a single postcursor tap ($h_1$); under these assumptions, $V_{\mathrm{IN}}[k]$ takes one of four noiseless levels, $\pm h_0 \pm h_1$, and we further assume $\max_i |h_i| = |h_0|$. The PAM-4 ($M=4$) instantiation is developed in Section~\ref{sec3:subsec1}.

We define the voltage error metric $e_{s,t}[k]$ as the distance between the received sample $V_{\mathrm{IN}}[k]$ and the noiseless level corresponding to a hypothesized previous bit $a[k-1]=s$ and current bit $a[k]=t$:
\begin{equation}
e_{s,t}[k] \eqdef h_{1}(2s-1) + h_0(2t-1) - V_{\mathrm{IN}}[k]
\end{equation}
where $s, t \in \{0,1\}$. This relationship is visualized in the trellis diagram of Fig.~\ref{fig:trellis}.

To perform Maximum-Likelihood (ML) detection over a two-UI window (Win-2), the accumulated squared Euclidean distance across two consecutive samples is minimized. For each candidate 3-bit sequence, the metric is defined as
\begin{equation}\label{eq:euclidean_dist1}
\begin{alignedat}{3}
a[k-2:k]&=[0,0,0]&&\leftrightarrow e_{0,0}^2[k-1]&&+e_{0,0}^2[k], \\ 
a[k-2:k]&=[0,0,1]&&\leftrightarrow e_{0,0}^2[k-1]&&+e_{0,1}^2[k], \\ 
a[k-2:k]&=[0,1,0]&&\leftrightarrow e_{0,1}^2[k-1]&&+e_{1,0}^2[k], \\ 
&&\vdots \\ 
a[k-2:k]&=[1,1,1]&&\leftrightarrow e_{1,1}^2[k-1]&&+e_{1,1}^2[k].
\end{alignedat}
\end{equation}
Although \eqref{eq:euclidean_dist1} enumerates all candidate paths, direct evaluation of every sequence is unnecessary. By regrouping terms with common subexpressions, the minimum accumulated distance can be written as
\begin{equation}
\min \begin{cases}
\begin{alignedat}{3}
&e_{0,0}^2[k-1]&&+\min(e_{0,0}^2[k],\ &&e_{0,1}^2[k]) \\ 
&e_{0,1}^2[k-1]&&+\min(e_{1,0}^2[k], &&e_{1,1}^2[k]) \\ 
&e_{1,0}^2[k-1]&&+\min(e_{0,0}^2[k], &&e_{0,1}^2[k]) \\ 
&e_{1,1}^2[k-1]&&+\min(e_{1,0}^2[k], &&e_{1,1}^2[k]).
\end{alignedat}
\end{cases}
\end{equation}
A further rearrangement yields the hardware-oriented form
\begin{equation} \label{eq:errorstatus_rearranged1}
\min\!\left\{
\begin{aligned}
&\min(e_{0,0}^2[k\mms1], e_{1,0}^2[k\mms1])\mps\min(e_{0,0}^2[k], e_{0,1}^2[k]) \\ 
&\min(e_{0,1}^2[k\mms1], e_{1,1}^2[k\mms1])\mps\min(e_{1,0}^2[k], e_{1,1}^2[k]).
\end{aligned}
\right.
\end{equation}
Equation \eqref{eq:errorstatus_rearranged1} reduces the Win-2 ML detector to a comparison between two composite error metrics. Each composite metric contains pairwise minimum operations evaluated at time indices $k-1$ and $k$. Importantly, each pairwise minimum can be selected by a simple threshold test on the received voltage.

For the terms at time index $k-1$, the selection depends on whether $V_{\mathrm{IN}}[k-1]$ is closer to $-h_0$ or $+h_0$:
\begin{align}
&\min(e_{0,0}^2[k\mms1], e_{1,0}^2[k\mms1])\myeq\begin{cases}
      e_{0,0}^2[k\mms1],\text{ if } V_{\mathrm{IN}}[k\mms1]\myle\mms h_0 \\ 
      e_{1,0}^2[k\mms1],\text{ otherwise.}
  \end{cases} \label{eq:errorstatus_km1} \\ 
&\min(e_{0,1}^2[k\mms1], e_{1,1}^2[k\mms1])\myeq\begin{cases}
      e_{1,1}^2[k\mms1],\text{ if } V_{\mathrm{IN}}[k\mms1]\mygeq h_0.\\
      e_{0,1}^2[k\mms1],\text{ otherwise. }
  \end{cases} \label{eq:errorstatus_km2}
\end{align}
According to \eqref{eq:errorstatus_rearranged1}, evaluating the minimum between \eqref{eq:errorstatus_km1} and \eqref{eq:errorstatus_km2} is still required. However, as shown later in this section, this comparison can be omitted.

Similarly, for the terms at time index $k$, the selection is determined by $h_1$:
\begin{align} 
&\min(e_{0,0}^2[k], e_{0,1}^2[k])=\begin{cases}
      e_{0,0}^2[k], & \text{if } V_{\mathrm{IN}}[k]\myle \mms h_1 \\ 
      e_{0,1}^2[k], & \text{otherwise.}
  \end{cases} \label{eq:errorstatus_k1}\\ 
&\min(e_{1,0}^2[k], e_{1,1}^2[k])=\begin{cases}
      e_{1,1}^2[k], & \text{if } V_{\mathrm{IN}}[k]\mygeq h_1. \\ 
      e_{1,0}^2[k], & \text{otherwise.} 
  \end{cases} \label{eq:errorstatus_k2}
\end{align}
The same threshold-based reasoning used in \eqref{eq:errorstatus_km1} and \eqref{eq:errorstatus_km2} directly applies to \eqref{eq:errorstatus_k1} and \eqref{eq:errorstatus_k2}.

Applying \eqref{eq:errorstatus_km1}--\eqref{eq:errorstatus_k2} partitions the $(V_{\mathrm{IN}}[k-1],V_{\mathrm{IN}}[k])$ plane into regions in which the current-bit decision is either immediately determined or remains ambiguous. In all regions except the central strip $-h_1<V_{\mathrm{IN}}[k]<h_1$, the competing paths may differ in their predecessor symbols but already agree on $\hat{a}[k]$, so the metric comparison can be omitted.

The Win-2 FFNE differs from Viterbi detection by restricting aggregation to a single 2-UI window and emitting only the current-bit decision $\hat{a}[k]$. This simplification is justified when the main cursor dominates the postcursor, so that predecessor-symbol ambiguity has limited impact on the current-bit decision.

For example, when $V_{\mathrm{IN}}[k-1] < -h_0$ and $V_{\mathrm{IN}}[k] < -h_1$, the composite metrics reduce to
\begin{equation}\label{eq:ffne_example1}
\begin{alignedat}{2}
&\min(e_{0, 0}^2[k\mms1],e_{1,0}^2[k\mms1])+\min(e_{0, 0}^2[k],e_{0,1}^2[k])\\
&\ \ =e_{0,0}^2[k\mms1]+e_{0,0}^2[k]\leftrightarrow \vec{\hat{a}}=[0,0,0],\\ 
&\min(e_{0, 1}^2[k\mms1],e_{1,1}^2[k\mms1])+\min(e_{1,0}^2[k],e_{1,1}^2[k])\\
&\ \ =e_{0,1}^2[k\mms1]+e_{1,0}^2[k]\leftrightarrow \vec{\hat{a}}=[0,1,0].\\ 
\end{alignedat}
\end{equation}
Although the two candidate sequences differ in their previous bits, both yield the same current-bit decision, $\hat{a}[k]=0$. Hence, the comparison between the two metrics is unnecessary.

Likewise, for $-h_0\leq V_{\mathrm{IN}}[k-1]\leq h_0$ and $V_{\mathrm{IN}}[k]<-h_1$, the candidate metrics become
\begin{equation} \label{eq:eq_11}
\begin{alignedat}{2}
&\min(e_{0, 0}^2[k\mms1],e_{1,0}^2[k\mms1])+\min(e_{0, 0}^2[k],e_{0,1}^2[k])\\
&\ \ =e_{1,0}^2[k\mms1]+e_{0,0}^2[k]\leftrightarrow \vec{\hat{a}}=[1,0,0],\\ 
&\min(e_{0, 1}^2[k\mms1],e_{1,1}^2[k\mms1])+\min(e_{1,0}^2[k],e_{1,1}^2[k])\\
&\ \ =e_{0,1}^2[k\mms1]+e_{1,0}^2[k]\leftrightarrow \vec{\hat{a}}=[0,1,0].\\ 
\end{alignedat}
\end{equation}

For $V_{\mathrm{IN}}[k-1]>h_0$ and $V_{\mathrm{IN}}[k]>h_1$, they become
\begin{equation} \label{eq:eq_12}
\begin{alignedat}{2}
&\min(e_{0, 0}^2[k\mms1],e_{1,0}^2[k\mms1])+\min(e_{0, 0}^2[k],e_{0,1}^2[k])\\
&\ \ =e_{1,0}^2[k\mms1]+e_{0,1}^2[k]\leftrightarrow \vec{\hat{a}}=[1,0,1],\\ 
&\min(e_{0, 1}^2[k\mms1],e_{1,1}^2[k\mms1])+\min(e_{1,0}^2[k],e_{1,1}^2[k])\\
&\ \ =e_{1,1}^2[k\mms1]+e_{1,1}^2[k]\leftrightarrow \vec{\hat{a}}=[1,1,1].\\ 
\end{alignedat}
\end{equation}

\begin{figure*}[!t]
    \centering
    \includegraphics[width=0.95\linewidth]{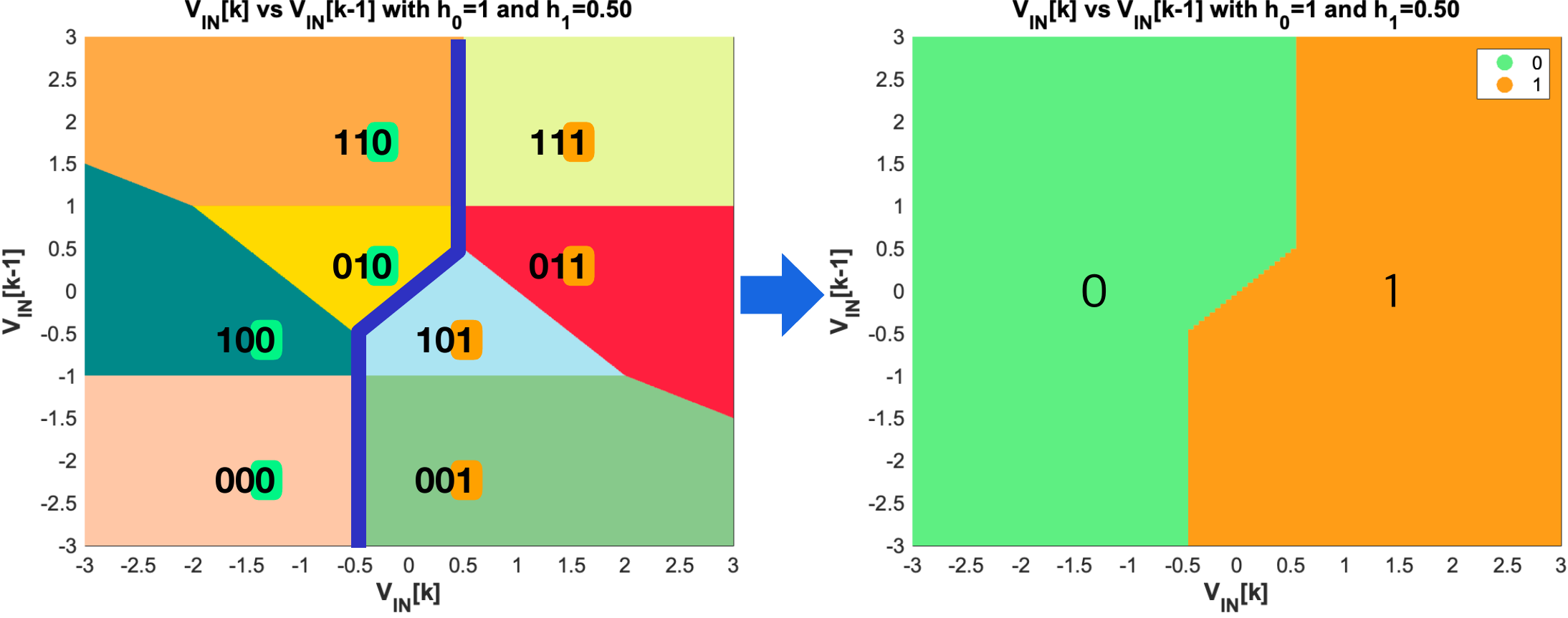}
    \caption{A key implementation simplification: by grouping the maximum-likelihood (ML) sequence regions according to the last bit, the multi-region dissection reduces to a compact binary decision rule. The labels in (a) denote the transmitted 3-bit sequence associated with each corresponding minimum-distance decision region.}
    \label{fig:ffmlse_mmse_map}
\end{figure*}

For $-h_0\leq V_{\mathrm{IN}}[k-1]\leq h_0$ and $V_{\mathrm{IN}}[k]>h_1$, they become
\begin{equation} \label{eq:eq_13}
\begin{alignedat}{2}
&\min(e_{0, 0}^2[k\mms1],e_{1,0}^2[k\mms1])+\min(e_{0, 0}^2[k],e_{0,1}^2[k])\\
&\ \ =e_{1,0}^2[k\mms1]+e_{0,1}^2[k]\leftrightarrow \vec{\hat{a}}=[1,0,1],\\ 
&\min(e_{0, 1}^2[k\mms1],e_{1,1}^2[k\mms1])+\min(e_{1,0}^2[k],e_{1,1}^2[k])\\
&\ \ =e_{0,1}^2[k\mms1]+e_{1,1}^2[k]\leftrightarrow \vec{\hat{a}}=[0,1,1].\\ 
\end{alignedat}
\end{equation}
Equations \eqref{eq:eq_11}--\eqref{eq:eq_13} show the same pattern: although the candidate paths differ in their predecessor symbols, they produce the same current-bit decision. The metric comparison is therefore unnecessary in these regions.

The remaining nontrivial case occurs when $-h_1<V_{\mathrm{IN}}[k]<h_1$. 
For the case of $-h_0\leq V_{\mathrm{IN}}[k-1]\leq h_0 \text{ and }-h_1<V_{\mathrm{IN}}[k]<h_1$, where the two candidate metrics are
\begin{equation} \label{eq:eq_14}
\begin{alignedat}{2}
&\min(e_{0, 0}^2[k\mms1],e_{1,0}^2[k\mms1])+\min(e_{0, 0}^2[k],e_{0,1}^2[k])\\
&\ \ =e_{1,0}^2[k\mms1]+e_{0,1}^2[k]\leftrightarrow \vec{\hat{a}}=[1,0,1],\\ 
&\min(e_{0, 1}^2[k\mms1],e_{1,1}^2[k\mms1])+\min(e_{1,0}^2[k],e_{1,1}^2[k])\\
&\ \ =e_{0,1}^2[k\mms1]+e_{1,0}^2[k]\leftrightarrow \vec{\hat{a}}=[0,1,0].
\end{alignedat}
\end{equation}
Unlike the previous cases, the current-bit decision now depends on which metric is smaller. The detector must therefore explicitly resolve
\begin{equation}\label{eq:middle_decision}
e_{1,0}^2[k\mms1]+e_{0,1}^2[k]\stackrel{?}{\gtrless}e_{0,1}^2[k\mms1]+e_{1,0}^2[k].
\end{equation}
The left-hand side of the inequality can be rewritten as 
\begin{equation}
\begin{alignedat}{1}
e_{1,0}^2[k\mms1]+e_{0,1}^2[k]&=(V_{\mathrm{IN}}[k-1]-(-h_0+h_1))^2\\&\phantom{=}+(V_{\mathrm{IN}}[k]-(h_0-h_1))^2.
\end{alignedat}
\end{equation}
The right-hand side of the inequality can be rewritten as 
\begin{equation}
\begin{alignedat}{1}
e_{0,1}^2[k-1]+e_{1,0}^2[k]&=(V_{\mathrm{IN}}[k-1]-(h_0-h_1))^2\\&\phantom{=}+(V_{\mathrm{IN}}[k]-(-h_0+h_1))^2.
\end{alignedat}
\end{equation}
Therefore, \eqref{eq:middle_decision} reduces to the comparison
\begin{equation}
V_{\mathrm{IN}}[k]\gtrless V_{\mathrm{IN}}[k-1],
\end{equation}
which leads to the decision rule
\begin{equation}\label{eq:win2ffne_center_decision}
\hat{a}[k-2:k]=\begin{cases}
    [1,0,1], & \text{if } V_{\mathrm{IN}}[k-1]\geq V_{\mathrm{IN}}[k],\\
    [0,1,0], & \text{otherwise.}
    \end{cases}
\end{equation}
Thus, explicit metric comparison is required only within this narrow central region. This is the key implementation simplification: the detector resolves an additional comparison only where the final-bit decision remains ambiguous. As illustrated in Fig. \ref{fig:ffmlse_mmse_map}, grouping the ML regions according to the last detected bit collapses the full sequence partition into a compact binary decision rule.

For the region $V_{\mathrm{IN}}[k-1] > h_0$ and $-h_1 < V_{\mathrm{IN}}[k] < h_1$, the candidate composite metrics reduce to
\begin{equation}
\begin{alignedat}{2}
&\min(e_{0, 0}^2[k\mms1],e_{1,0}^2[k\mms1])+\min(e_{0, 0}^2[k],e_{0,1}^2[k])\\
&\ \ =e_{1,0}^2[k\mms1]+e_{0,1}^2[k]\leftrightarrow \hat{a}[k]=[1,0,1],\\ 
&\min(e_{0, 1}^2[k\mms1],e_{1,1}^2[k\mms1])+\min(e_{1,0}^2[k],e_{1,1}^2[k])\\
&\ \ =e_{1,1}^2[k\mms1]+e_{1,0}^2[k]\leftrightarrow \hat{a}[k]=[1,1,0].
\end{alignedat}
\end{equation}
The comparison between these two candidates reduces to
\begin{equation}
\begin{alignedat}{2}
e_{1,0}^2[k\mms1]+e_{0,1}^2[k] &\stackrel{?}{\gtrless}e_{1,1}^2[k\mms1]+e_{1,0}^2[k] \\ 
e_{1,0}^2[k\mms1]+e_{0,1}^2[k]&=(V_{\mathrm{IN}}[k-1]-(-h_0+h_1))^2\\&\phantom{=}+(V_{\mathrm{IN}}[k]-(h_0-h_1))^2\\
e_{1,1}^2[k-1]+e_{1,0}^2[k]&=(V_{\mathrm{IN}}[k-1]-(h_0+h_1))^2\\&\phantom{=}+(V_{\mathrm{IN}}[k]-(-h_0+h_1))^2
\end{alignedat}
\end{equation}
which yields
\begin{equation}
\frac{V_{\mathrm{IN}}[k-1] - h_1}{1 - h_1/h_0} > V_{\mathrm{IN}}[k].
\end{equation}
To show that this inequality always holds over the region of interest, it is sufficient to evaluate the boundary values of $V_{\mathrm{IN}}[k-1] > h_0$ and $-h_1 < V_{\mathrm{IN}}[k] < h_1$:
\begin{equation}\label{eq:ambiguity_verification}
\begin{alignedat}{2}
&V_{\mathrm{IN}}[k\mms1]\myeq h_0,V_{\mathrm{IN}}[k]\myeq\mms h_1&&\rightarrow\frac{h_0-h_1}{1-h_1/h_0}>-h_1,\\ 
&V_{\mathrm{IN}}[k\mms1]\myeq \infty,V_{\mathrm{IN}}[k]\myeq\mms h_1&&\rightarrow\frac{\infty-h_1}{1-h_1/\infty}>-h_1,\\
&V_{\mathrm{IN}}[k\mms1]\myeq h_0,V_{\mathrm{IN}}[k]\myeq h_1&&\rightarrow\frac{h_0-h_1}{1-h_1/h_0}> h_1,\\
&V_{\mathrm{IN}}[k\mms1]\myeq \infty,V_{\mathrm{IN}}[k]\myeq  h_1&&\rightarrow\frac{\infty-h_1}{1-h_1/\infty}> h_1.
\end{alignedat}
\end{equation}
Therefore, throughout the region $V_{\mathrm{IN}}[k-1]>h_0 \text{ and } -h_1 < V_{\mathrm{IN}}[k] < h_1$, the following inequality always holds: 
\begin{equation}
e_{1,1}^2[k-1]+e_{1,0}^2[k] < e_{1,0}^2[k-1]+e_{0,1}^2[k],
\end{equation}
and the sequence [1,1,0] is always preferred over [1,0,1].

By symmetry, the same argument applies to the case $V_{\mathrm{IN}}[k-1]<-h_0$ and $-h_1<V_{\mathrm{IN}}[k]<h_1$. The resulting conditions for the central strip can therefore be summarized as follows:
\begin{enumerate}
    \item For $V_{\mathrm{IN}}[k-1]>h_0 \text{ and } -h_1<V_{\mathrm{IN}}[k]<h_1$:
    \begin{equation*}
        e_{1,1}^2[k-1]+e_{1,0}^2[k]<e_{1,0}^2[k-1]+e_{0,1}^2[k].
    \end{equation*}
    
    \item For $-h_0<V_{\mathrm{IN}}[k-1]<h_0 \text{ and } -h_1<V_{\mathrm{IN}}[k]<h_1$:
    \begin{equation*}
    \hspace*{-\leftmargin}
    \begin{cases}
        e_{1,0}^2[k\mms1]\mps e_{0,1}^2[k] \myleq e_{0,1}^2[k\mms 1]\mps e_{1,0}^2[k], & \text{if }V_{\mathrm{IN}}[k\mms 1]\myle V_{\mathrm{IN}}[k] \\ 
        e_{1,0}^2[k\mms1]\mps e_{0,1}^2[k] \myge e_{0,1}^2[k\mms 1]\mps e_{1,0}^2[k], & \text{otherwise} 
    \end{cases}
    \end{equation*}
    
    \item For $V_{\mathrm{IN}}[k-1]<-h_0 \text{ and } -h_1<V_{\mathrm{IN}}[k]<h_1$:
    \begin{equation*}
        e_{0,0}^2[k-1]+e_{0,1}^2[k]<e_{0,1}^2[k-1]+e_{1,0}^2[k].
    \end{equation*}
\end{enumerate}

\begin{figure}[!t]
    \centering
    \includegraphics[width=0.95\columnwidth]{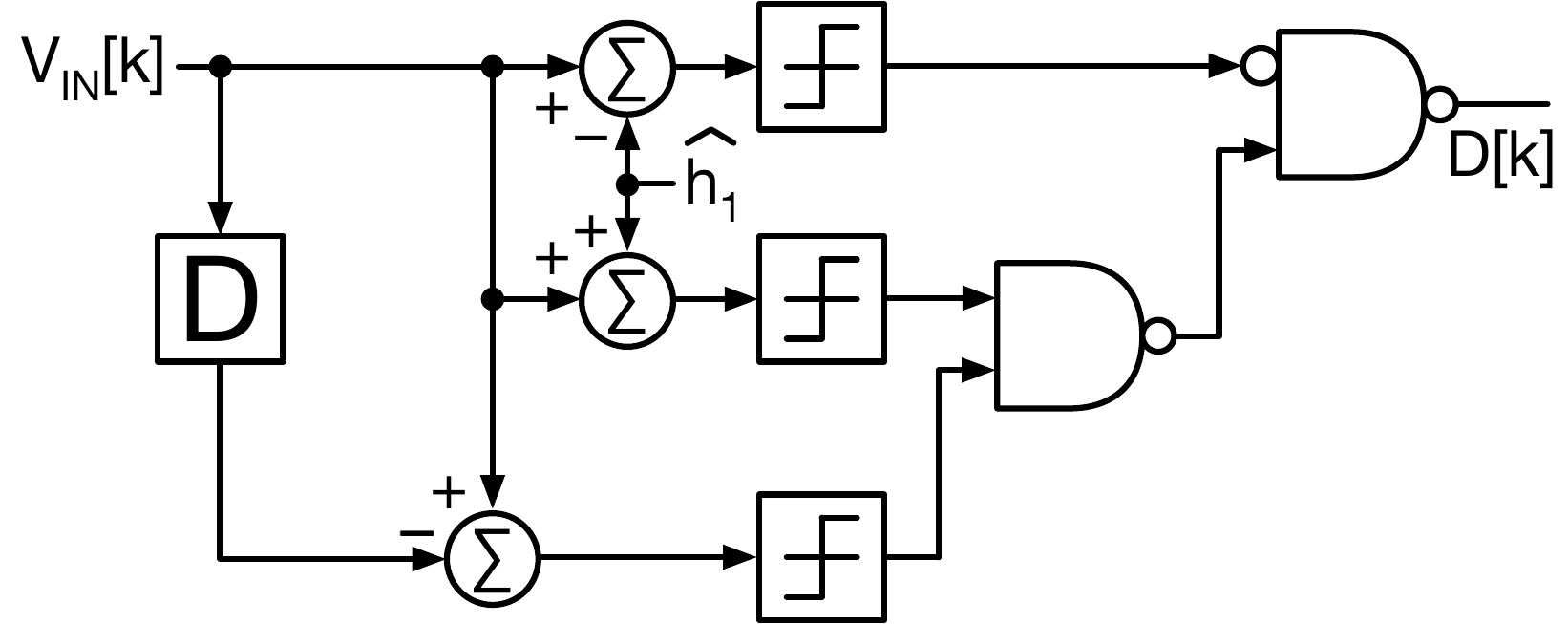}
    \caption{Hardware implementation of the Win-2 FFNE for $h_1$ cancellation in NRZ. `D' denotes an analog 1-UI delay element.}
    \label{fig:win2ffmlse}
\end{figure}

The derivation above admits a natural geometric interpretation, shown in Fig. \ref{fig:ffmlse_mmse_map}. For a channel with main cursor $h_0$ and single postcursor $h_1$, each 3-symbol transmit sequence $\vec{a}$ maps to a unique noiseless point in the $(V_{\mathrm{IN}}[k],V_{\mathrm{IN}}[k-1])$ plane. Under AWGN, the ML detector selects the sequence with minimum Euclidean distance, thereby partitioning the plane into Voronoi-like decision regions.

The key implementation step in this work is to modify these boundaries for hardware efficiency by grouping regions according to the last detected bit rather than preserving the full ideal Voronoi partition. While the resulting geometry remains visually reminiscent of prior work \cite{odling_thesis, emami_ffne}, the grouped partition collapses the original multi-region ML map into a compact binary decision map. As a result, the grouped piecewise boundaries can be implemented using only three comparators and two NAND gates, yielding the high-speed, low-power Win-2 FFNE shown in Fig.~\ref{fig:win2ffmlse} without the timing overhead of feedback-based equalization.

\subsection{Tap-Coefficients Adaptation} \label{sec2:subsec2}

\begin{figure}[!t]
    \centering
    \includegraphics[width=0.95\columnwidth]{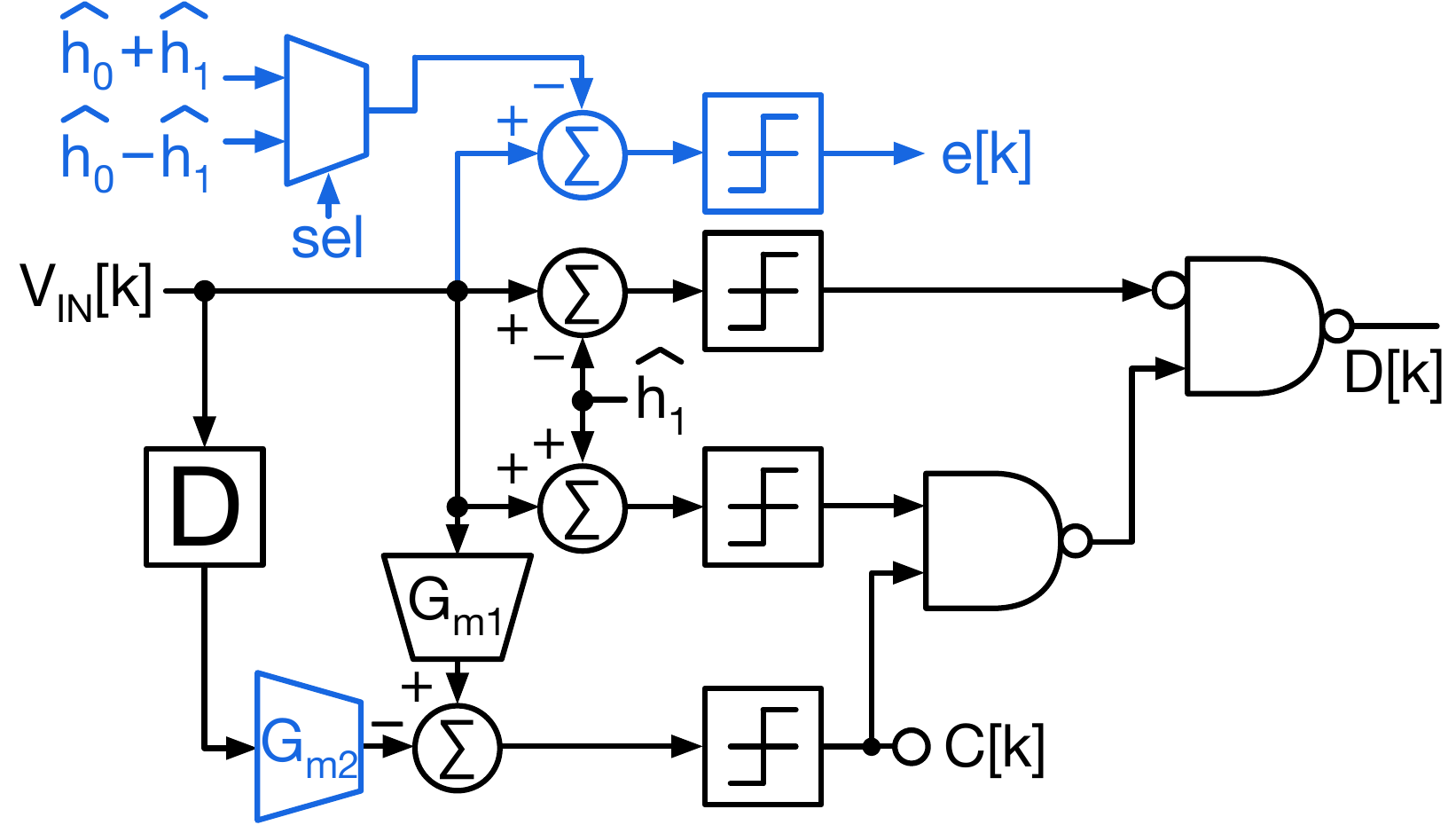}
    \caption{Error slicer implementation for $\hat{h}_1$ adaptation and summer implementation in Win-2 FFNE.}
    \label{fig:win2mlse_adaptation}
\end{figure}

\begin{figure}
    \centering
    \captionsetup[subfigure]{justification=centering}
        \begin{subfigure}[b]{\columnwidth}
        \centering
        \includegraphics[width=0.91\columnwidth]{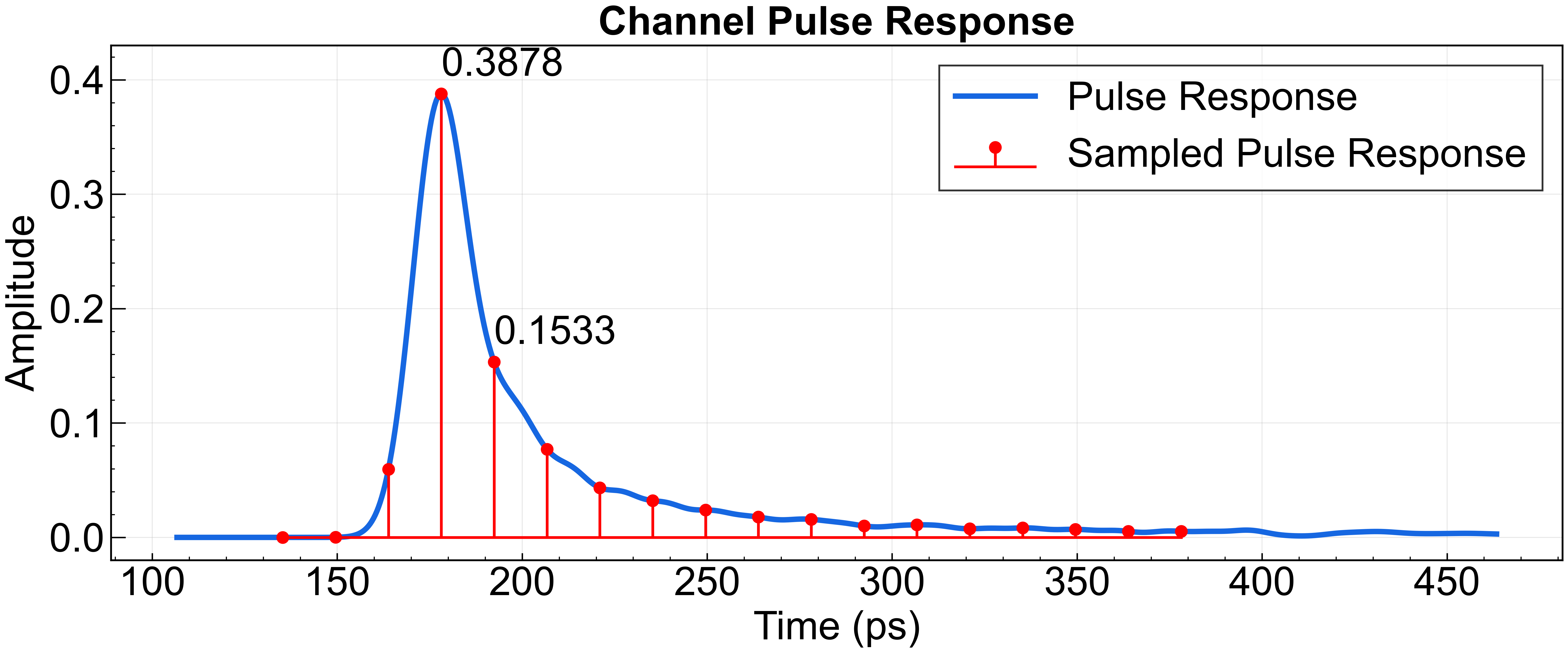}
        \caption{}
        \end{subfigure}
        
        \hfill
        \begin{subfigure}[b]{\columnwidth}
        \centering
        \includegraphics[width=0.91\columnwidth]{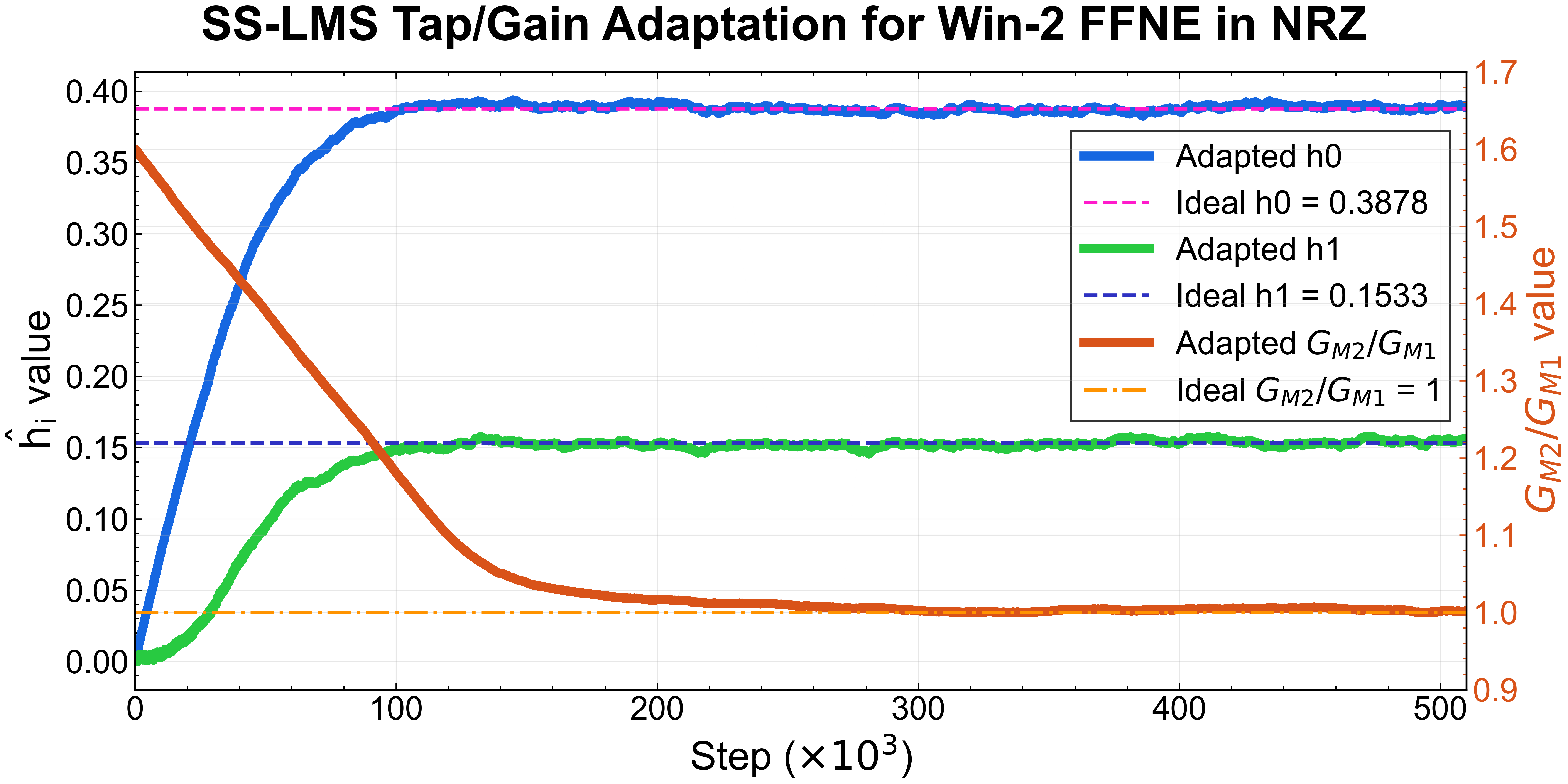}
        \caption{}
        \end{subfigure}

    \caption{(a) Pulse response with a channel insertion loss of 15 dB at Nyquist frequency, and (b) adapted tap values using the SS-LMS.}
    \label{fig:win2mlse_adaptation_plot}
\end{figure}

The architectural similarity between the Win-2 FFNE and the 1-tap loop-unrolled DFE allows $dLev$-based sign-sign LMS (SS-LMS) adaptation \cite{vlad_adaptation}. Slicer inputs ideally follow the four-level pattern $\pm h_0 \pm h_1$, where $h_0 \pm h_1$ are derived from the 11 and 01 data patterns. With $\hat{h}_i^k$ representing the adapted estimate of $h_i$, the update rules are:

\begin{equation}\label{eq:win2ffne_adaptation1}\
\begin{alignedat}{1}
    dLev_{11}^{k} &= dLev_{11}^{k-1} + \mu\, e[k]\, D[k] \\ 
    & \text{ update when }D[k-1]=1\wedge D[k]=1, \\ 
    dLev_{01}^{k} &= dLev_{01}^{k-1} + \mu\, e[k]\, D[k] \\ 
    & \text{ update when }D[k-1]=0\wedge D[k]=1 
\end{alignedat}
\end{equation}

from which the tap estimates can be obtained as
\begin{equation}\label{eq:win2ffne_adaptation2}
\hat{h}_0^k = \frac{dLev_{11}^k + dLev_{01}^k}{2},
\qquad
\hat{h}_1^k = \frac{dLev_{11}^k - dLev_{01}^k}{2}.
\end{equation}

Beyond $h_0$ and $h_1$ adaptation, accurate Win-2 FFNE operation also requires matched gain in the summer that evaluates $V_{\mathrm{IN}}[k]-V_{\mathrm{IN}}[k-1]$. In a current-summing implementation, mismatch between $G_{m1}$ and $G_{m2}$ can be adapted using repeated-symbol patterns, for which the ideal summer output vanishes. Specifically, for the 111 and 000 patterns,

\begin{figure}[!t]
    \centering
    \includegraphics[width=0.95\columnwidth]{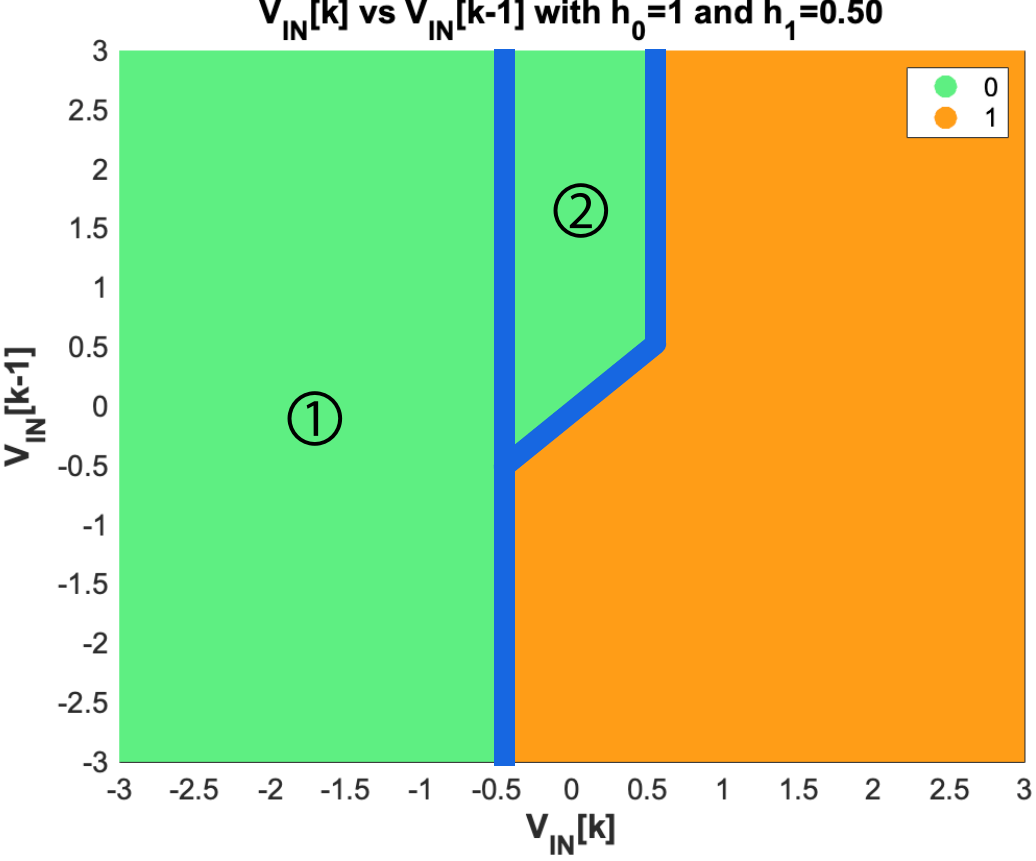}
    \caption{Geometric decomposition of window-length-2 FFNE decision boundaries for closed-form statistical analysis.}
    \label{fig:win2mlse_stat}
\end{figure}

\begin{figure}[!t]
    \centering
    \includegraphics[width=0.95\columnwidth]{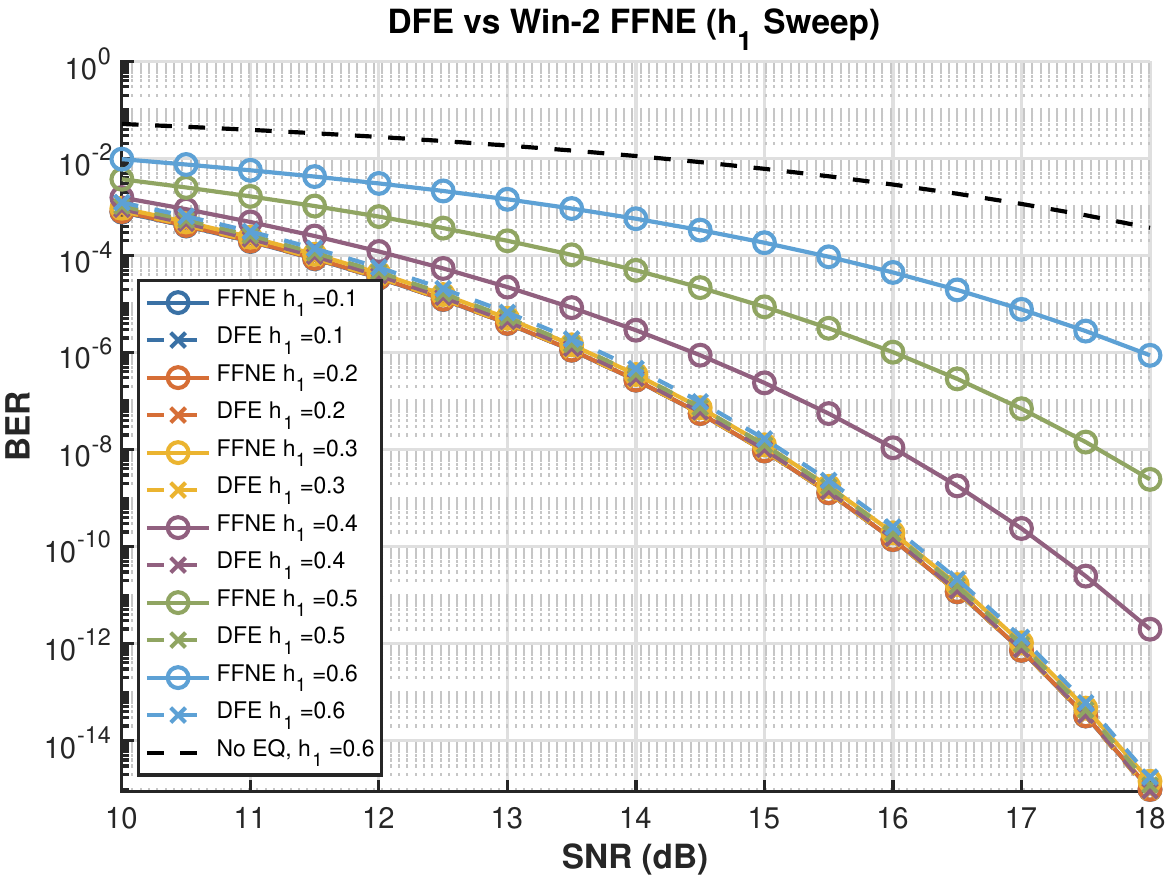}
    \caption{Statistical-model-based BER comparison of the NRZ 1-tap DFE and the Win-2 FFNE for $h_1$ values from 0.1 to 0.6.}
    \label{fig:dfe_vs_ffne}
\end{figure}

\begin{equation}\label{eq:summer_error}
\begin{alignedat}{1}
V_{\Sigma,111} &= (h_0+h_1)G_{m1}-(h_0+h_1)G_{m2}, \\
V_{\Sigma,000} &= (-h_0-h_1)G_{m1}-(-h_0-h_1)G_{m2}.
\end{alignedat}
\end{equation}

If the channel contains only the two ISI components, both expressions vanish when $G_{m1}=G_{m2}$, and any residual output provides an error polarity for gain adaptation. The slicer output $C[k]$ is therefore used to update $G_{m2}$ as

\begin{equation}\label{eq:summer_sslms}
\begin{alignedat}{1}
\hat{G}_{m2}^{k} &= \hat{G}_{m2}^{k-1} + \mu C[k]D[k], \\
& \text{updated when } D[k-2]=D[k-1]=D[k].
\end{alignedat}
\end{equation}

An example hardware implementation is shown in Fig. \ref{fig:win2mlse_adaptation}. The error reference levels for generating \(e[k]\) can be time-division multiplexed to reduce the number of additional comparators, similar to \cite{lsi_25gbps}. 

Fig. \ref{fig:win2mlse_adaptation_plot} illustrates the pulse response of a channel with 15 dB insertion loss at the Nyquist frequency and adapted tap trajectories (\(\hat{h}_0^k\), \(\hat{h}_1^k\), and \(\hat{G}_m^k\)) obtained using \eqref{eq:win2ffne_adaptation1}--\eqref{eq:summer_sslms}. The convergence behavior shows that the proposed SS-LMS updates remain stable even with residual ISI components beyond the single-tap model, since the adaptation relies on pattern filtering.

\subsection{Statistical Analysis and EQ limitation} \label{sec2:subsec3}

\begin{figure*}[!t]
    \centering
    \includegraphics[width=0.95\linewidth]{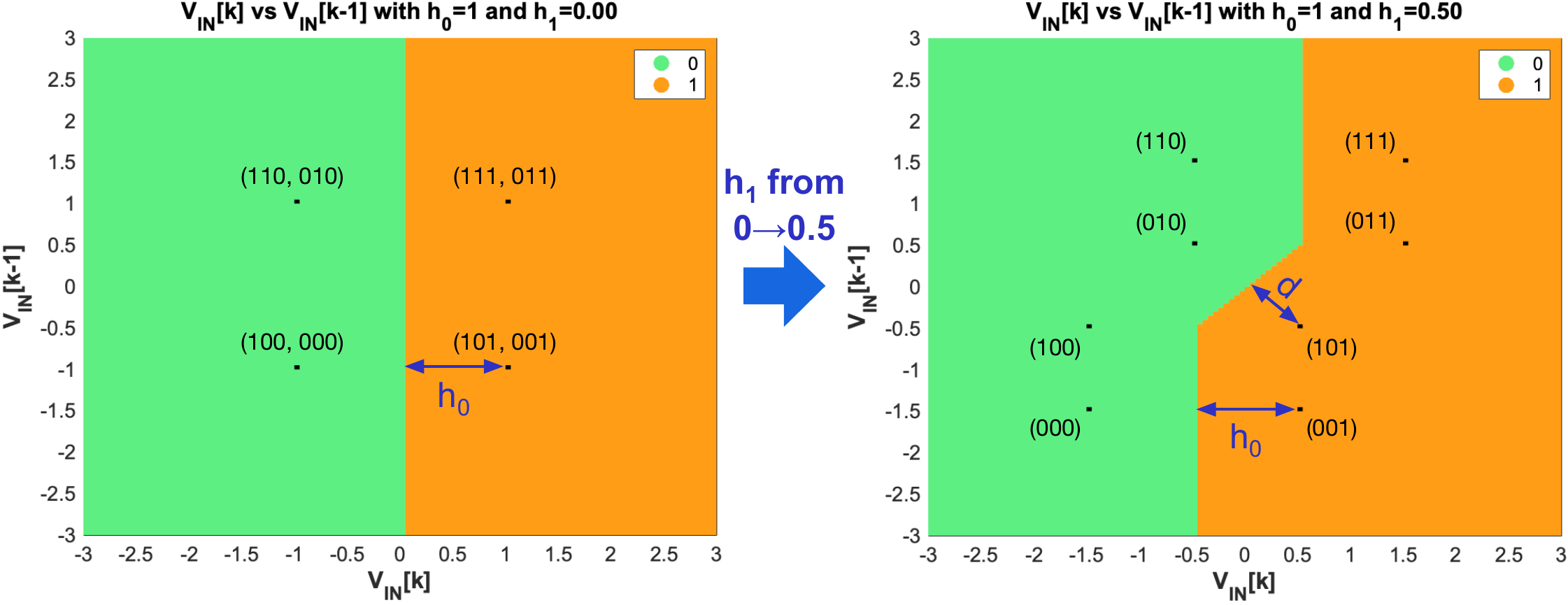}
    \caption{Migration of the window-length-2 FFNE noiseless constellation points in the 
    $(V_{\mathrm{IN}}[k], V_{\mathrm{IN}}[k-1])$ plane as $h_1$ is swept. The alternating sequences
    ([0,1,0] and [1,0,1]) move toward the origin and eventually dominate the error events.}
    \label{fig:mlse_limit_1}
\end{figure*}

For NRZ signaling, the error probability at time index $k$ is the conditional probability of deciding 0 when the transmitted bit is 1. Geometrically, this event comprises regions $\circled{1}$ and $\circled{2}$ in Fig. \ref{fig:win2mlse_stat}. Assuming each comparator has mutually uncorrelated noise, the combined effect is modeled as an AWGN term added to $V_{\mathrm{IN}}[k]$ with variance $\sigma_n^2$. Although the transmitted sequence is stationary, the DC offsets $\mu_{V_{\mathrm{IN}}[k]}$ and $\mu_{V_{\mathrm{IN}}[k-1]}$ may differ due to their distinct signal paths (Fig. \ref{fig:win2ffmlse}). With $a[k]=1$, the BER is:
\begin{equation}\label{eq_ber_circled}
    \begin{split}
    &BER=\underbrace{P(V_{\mathrm{IN}}[k]\leq -h_1)}_{\circled{1}}+\\
    &\underbrace{P(V_{\mathrm{IN}}[k]-V_{\mathrm{IN}}[k-1]\leq0, -h_1 \leq V_{\mathrm{IN}}[k] \leq h_1)}_{\circled{2}}.
    \end{split}
\end{equation}
The two probabilities in \eqref{eq_ber_circled} can be computed as follows.

\begin{align}
    & \circled{1}:P(V_{\mathrm{IN}}[k]\leq -h_1) = Q((\mu_{V_{\mathrm{IN}}[k]}+h_1)/\sigma_n), \\ 
    & \circled{2}:P(V_{\mathrm{IN}}[k]-V_{\mathrm{IN}}[k-1]\leq0, -h_1 \leq V_{\mathrm{IN}}[k] \leq h_1)=\nonumber \\ &\int_{-h_1}^{h_1}{\int_{V_{\mathrm{IN}}[k]}^{\infty}{\frac{1}{\sqrt{(2\pi)^k|\Sigma|}}e^{\left(-\frac{1}{2}\left[S^T\Sigma^{-1}S\right]\right)}}}dV_{\mathrm{IN}}[k-1]dV_{\mathrm{IN}}[k],
\end{align} where 
\begin{align}
    &S=\left[\begin{matrix}V_{\mathrm{IN}}[k]-\mu_{V_{\mathrm{IN}}[k]}\\V_{\mathrm{IN}}[k-1]-\mu_{V_{\mathrm{IN}}[k-1]} \end{matrix} \right],\\
    &\Sigma=\sigma_n^2\cdot\left[\begin{matrix}
                \sum_{i=1}^{J}{c_i^2} & \sum_{i=1}^{J-1}{(c_i\cdot c_{i+1})}\\
                \sum_{i=1}^{J-1}{(c_i\cdot c_{i+1})} & \sum_{i=1}^{J}{c_i^2}
                \end{matrix}
    \right],\\ 
    \quad &\vec{c}_{FFE}=\left[\begin{matrix} 
        c_1 & c_2 & \cdots & c_{J}
    \end{matrix}
    \right]. 
\end{align} 
Here, $\Sigma\in\mathbb{R}^{w\times w}$, where $w$ corresponds to the window length of FFNE (2 in this case), and $\vec{c}_{FFE}\in\mathbb{R}^{J}$, where $J$ corresponds to the number of FFE taps. 

Due to noise correlation from the Tx/Rx-FFE, we evaluate the error area $\circled{2}$ using a multivariate Gaussian CDF, where the FFE taps define the covariance matrix $\Sigma$ for the summer output $V_{\mathrm{IN}}[k]-V_{\mathrm{IN}}[k-1]$.

Fig. \ref{fig:dfe_vs_ffne} compares the BER of the Win-2 NRZ FFNE against a 1-tap DFE for postcursor values $h_1$ from 0.1 to 0.6 (assuming $h_0=1$). When $h_1 \leq 0.3$, both equalizers perform nearly identically. However, as $h_1$ increases beyond 0.3, the Win-2 FFNE exhibits a progressively larger BER penalty relative to the DFE. This is not a failure of equalization; even for large $h_1$, the FFNE still significantly outperforms an unequalized signal. Rather, the result indicates that its effective noise margin shrinks faster than that of an ideal DFE under strong ISI. The statistical-analysis source code is available in the repository listed in the footnote\footnote{The statistical analysis framework for this work is available at \url{https://github.com/kunmok/FFNE-stat-and-time-domain-simulator}.} 

This degradation stems from the sequence-space geometry. The Win-2 FFNE detects $a[k]$ by selecting the closest 3-bit sequence point in the $(V_{\mathrm{IN}}[k], V_{\mathrm{IN}}[k-1])$ plane. As shown in Fig. \ref{fig:mlse_limit_1}, increasing $h_1$ does not affect all sequence points uniformly. While most points remain safely distant from the decision threshold, the alternating patterns [0,1,0] and [1,0,1] are pulled toward the origin. Consequently, these patterns increasingly dominate the error probability as $h_1$ increases.  

To quantify this, consider the noiseless coordinates for the [1,0,1] sequence, $(a[k-2],a[k-1],a[k])=(1,0,1)$: 
\begin{equation}
\begin{alignedat}{1}
    V_{\mathrm{IN}}[k-1] &= -h_0 + h_1, \\
    V_{\mathrm{IN}}[k] &= +h_0 - h_1.
\end{alignedat}
\end{equation}
Its Euclidean distance from the origin ($d$) is therefore
\begin{equation}\label{eq:ffne_dmin}
    d \myeq \sqrt{(V_{\mathrm{IN}}[k\mms1])^2 \mps (V_{\mathrm{IN}}[k])^2} \myeq \sqrt{2h_0^2\mms4h_0h_1\mps2h_1^2}.
\end{equation}
The complementary alternating pattern [0,1,0] has the same distance. Hence, as \(h_1\) increases, these two points move toward the origin and increasingly dominate the error probability because they have the smallest noise margin.

We can identify a specific breakpoint where the Win-2 FFNE begins to underperform the DFE. An ideal 1-tap DFE (assuming no error propagation) maintains a constant decision margin of $h_0$ regardless of $h_1$. By setting our FFNE distance $d = h_0$ in \eqref{eq:ffne_dmin}, we find:
\begin{align}\label{eq:win2_h1_limit}
    h_1 &=h_0\cdot\left(1-\frac{1}{\sqrt{2}}\right)\approx h_0\cdot0.293.
\end{align}
Thus, once $h_1$ exceeds \eqref{eq:win2_h1_limit}, [0,1,0/1,0,1] sequences lie within a radius $h_0$ of the origin, making the FFNE more sensitive to noise than the DFE.

While increasing the FFNE window length could alleviate this limitation by improving the minimum distance of the sequence set, doing so would increase hardware complexity exponentially.

\subsection{Window-Length-3 FFNE for Enhanced EQ Capability} \label{sec2:subsec4}

\begin{figure*}[!t]
    \centering
    \includegraphics[width=0.95\linewidth]{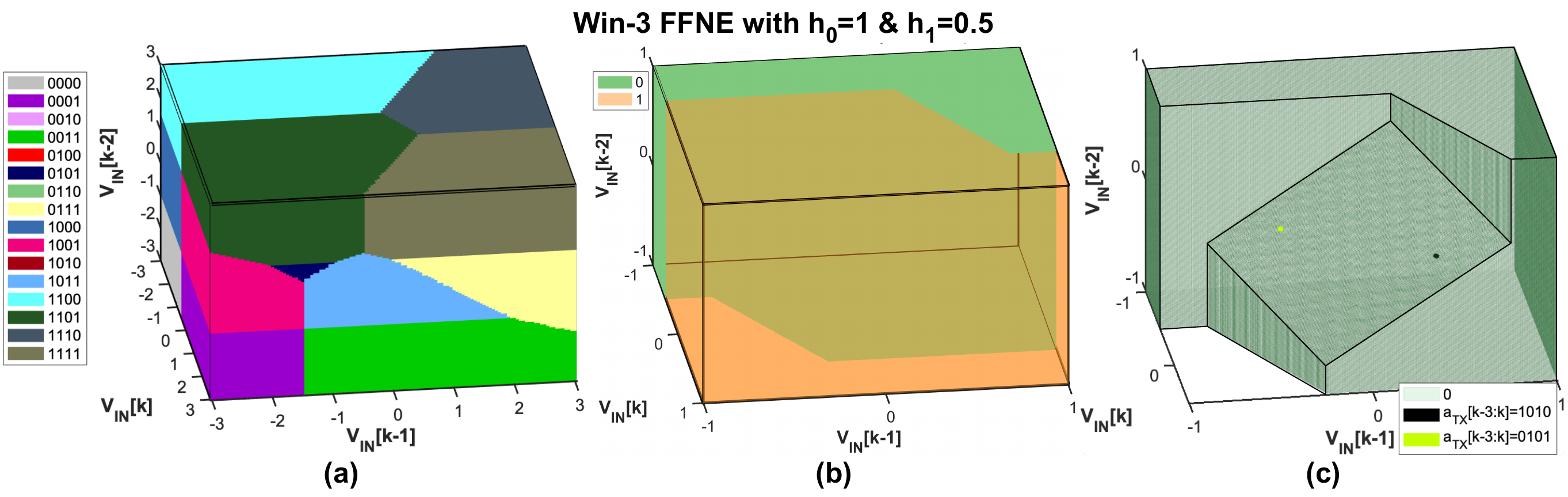}
    \caption{(a) Full 3-D Voronoi diagram of the window-length-3 FFNE. (b) Binary grouping of the same 3-D decision space according to the detected bit. (c) Cross-sectional view of the same geometry for improved visualization and intuition, with the yellow and black dots located at $(0.5,\,-0.5,\,0.5)$ and $(-0.5,\,0.5,\,-0.5)$, respectively, in the $(V_{\mathrm{IN}}[k],\,V_{\mathrm{IN}}[k\!-\!1],\,V_{\mathrm{IN}}[k\!-\!2])$ space.}
    \label{fig:win3_ffne_decision_cube}
\end{figure*}

\begin{figure*}[!t]
    \centering
    \includegraphics[width=0.95\linewidth]{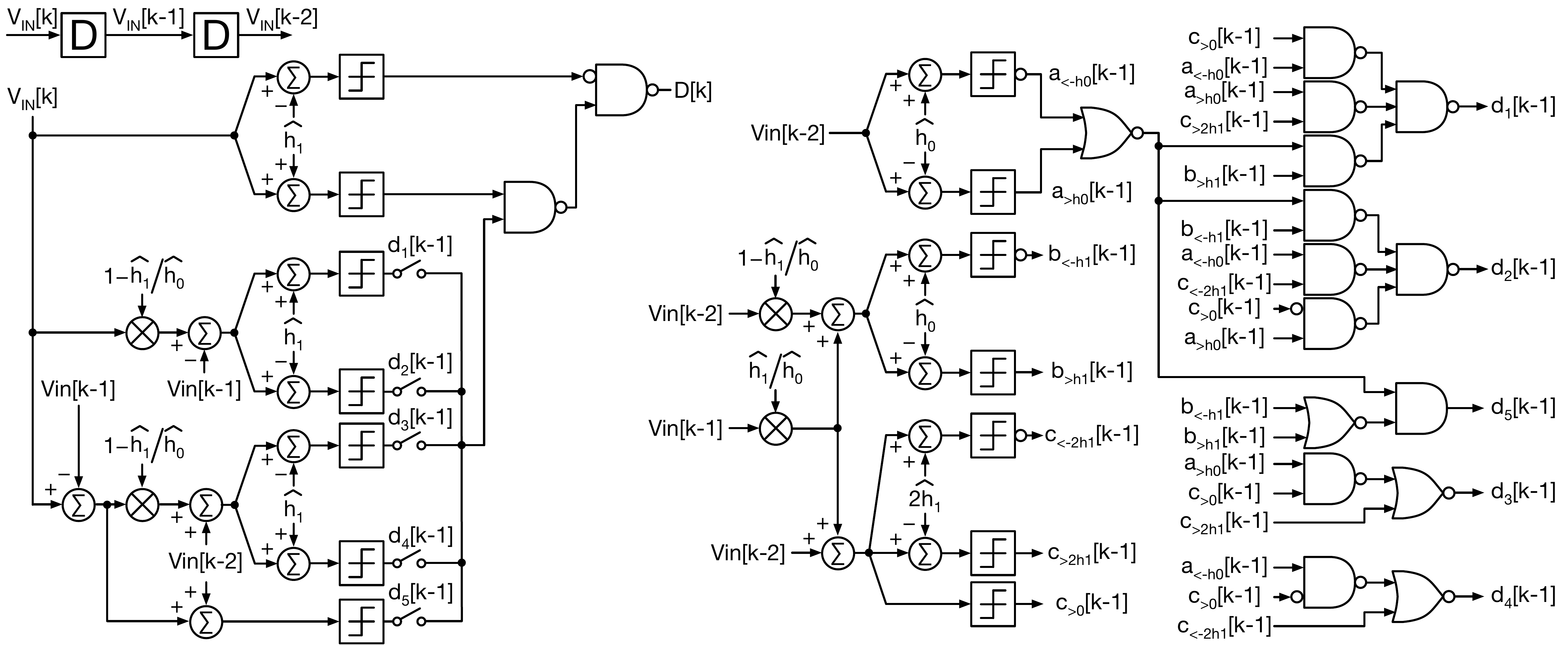}
    \caption{Hardware implementation of the window-length-3 FFNE for $h_1$ cancellation in NRZ modulation.}
    \label{fig:win3_diagram}
\end{figure*}

\begin{figure}[!t]
    \centering
    \includegraphics[width=0.95\columnwidth]{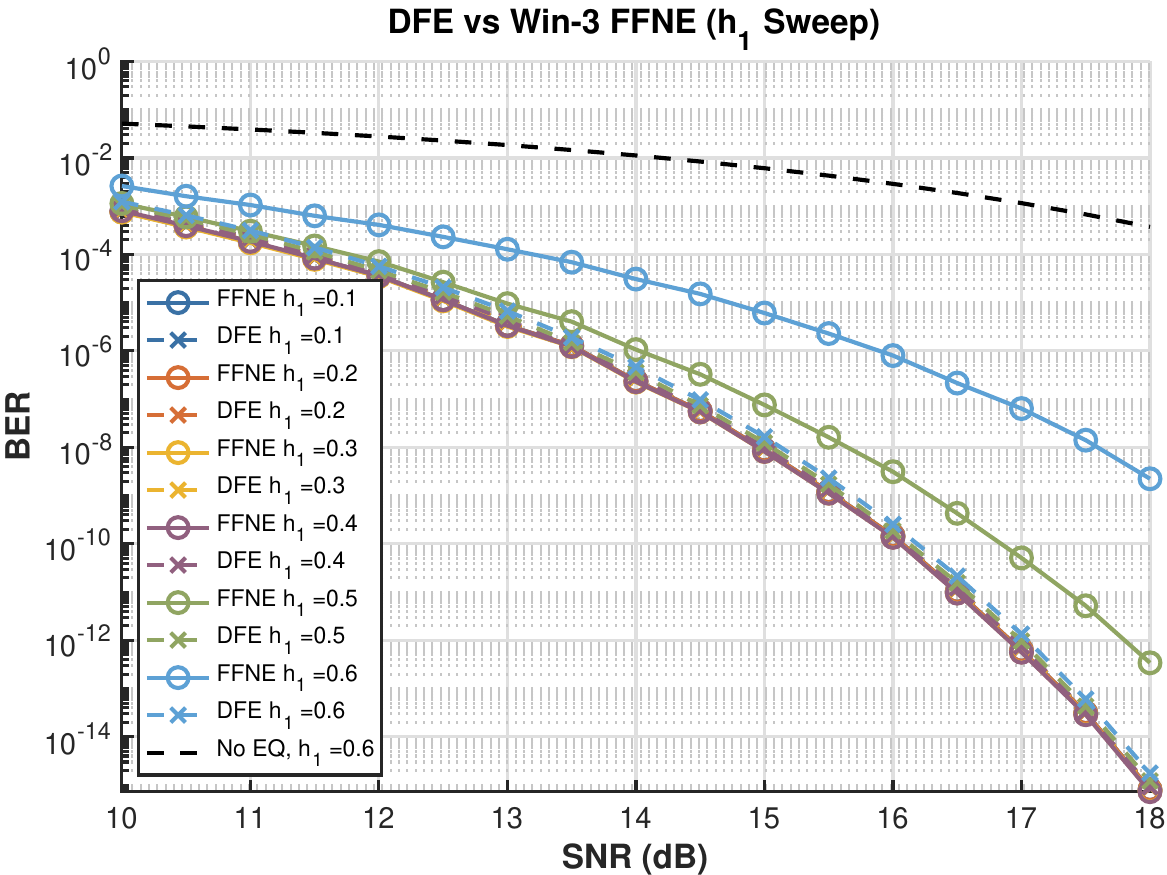}
    \caption{Statistical-model-based BER comparison of the NRZ 1-tap DFE and the window-length-3 FFNE for $h_1$ values from 0.1 to 0.6.}
    \label{fig:dfe_vs_ffne_win3}
\end{figure}
While the full mathematical derivation for the window-length-3 FFNE (Win-3 FFNE) is provided in the supplementary material, it follows the same fundamental principles as the Win-2 version. The primary distinction is that the Win-3 FFNE incorporates an additional observation point, $V_{\mathrm{IN}}[k-2]$, alongside $V_{\mathrm{IN}}[k-1]$ and $V_{\mathrm{IN}}[k]$.

\begin{figure*}
\centering
\begin{subfigure}{\columnwidth}
  \centering
  \includegraphics[width=0.95\columnwidth]{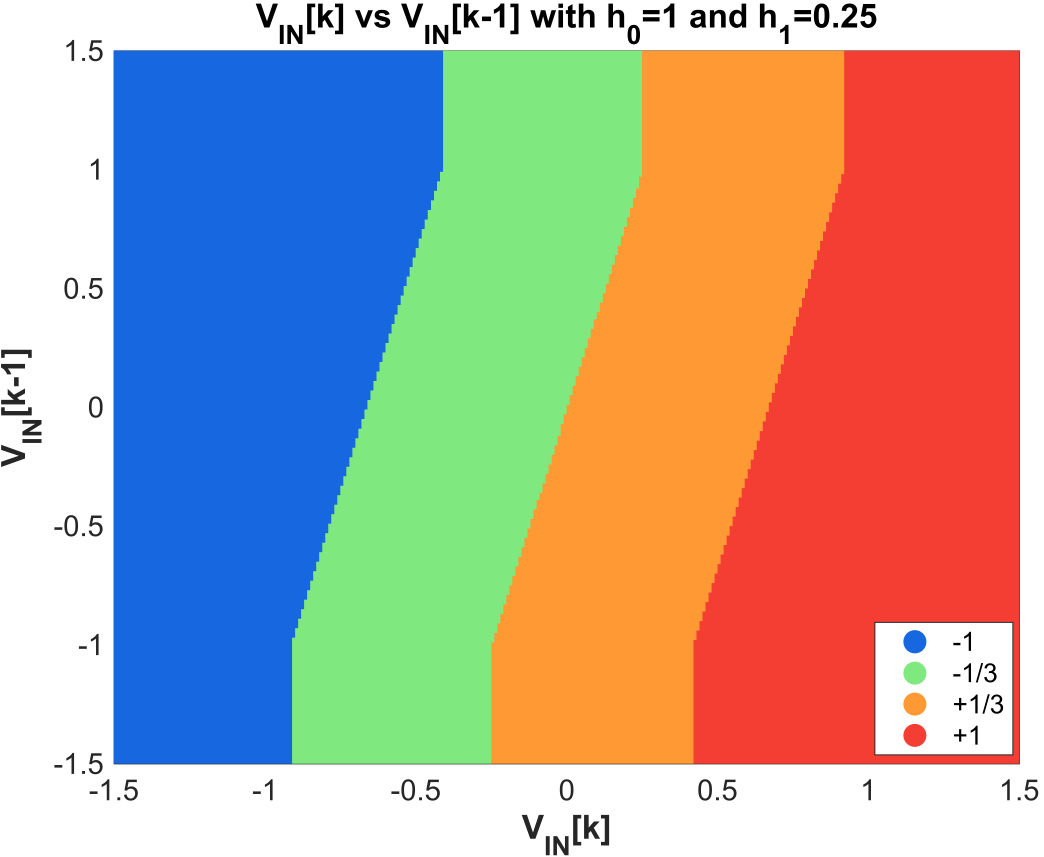}
  \caption{}
\end{subfigure}%
\begin{subfigure}{.5\textwidth}
  \centering
  \includegraphics[width=0.95\columnwidth]{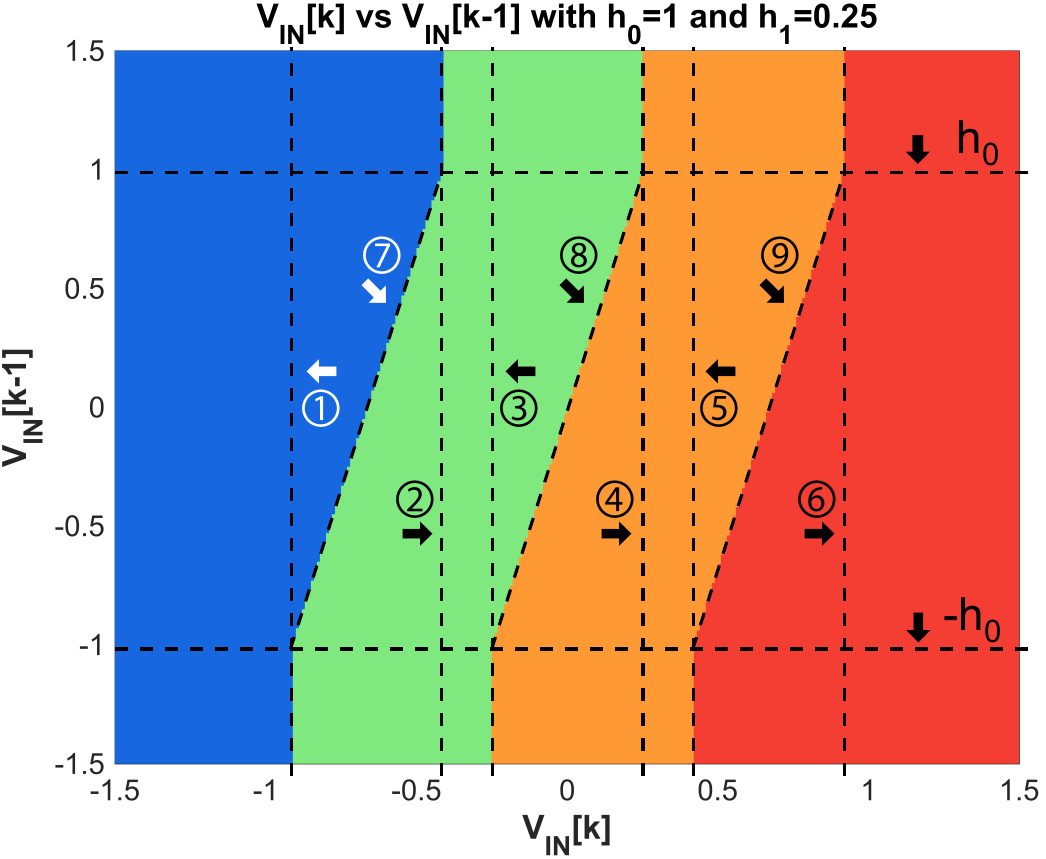}
  \caption{}
\end{subfigure}
\caption{(a) Decision regions of the PAM-4 window-length-2 FFNE in the $(V_{\mathrm{IN}}[k],V_{\mathrm{IN}}[k-1])$ plane, colored by the detected symbol $\hat{a}[k]\in\{\mms1,\mms\tfrac{1}{3},\tfrac{1}{3},1\}$. (b) Corresponding decision-boundary geometry used for the hardware implementation.}
\label{fig:ffmlse_pam4}
\end{figure*}

\begin{figure}[!t]
    \centering
    \includegraphics[width=0.95\columnwidth]{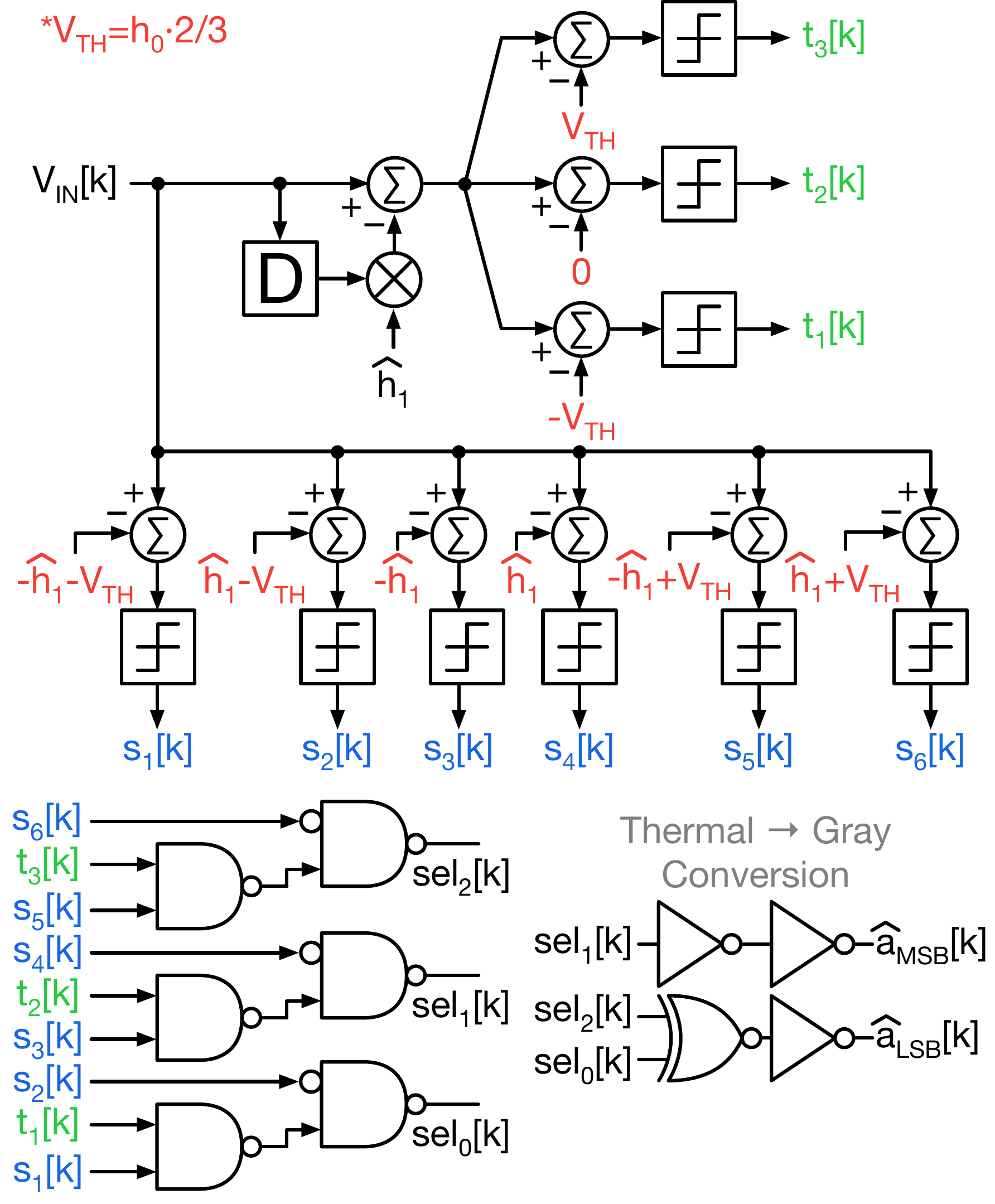}
    \caption{Hardware implementation of the Win-2 FFNE for $h_1$ cancellation in PAM-4 modulation.}
    \label{fig:pam4_ffne_diagram}
\end{figure}

\begin{figure}[ht]
    \centering
    \includegraphics[width=0.95\columnwidth]{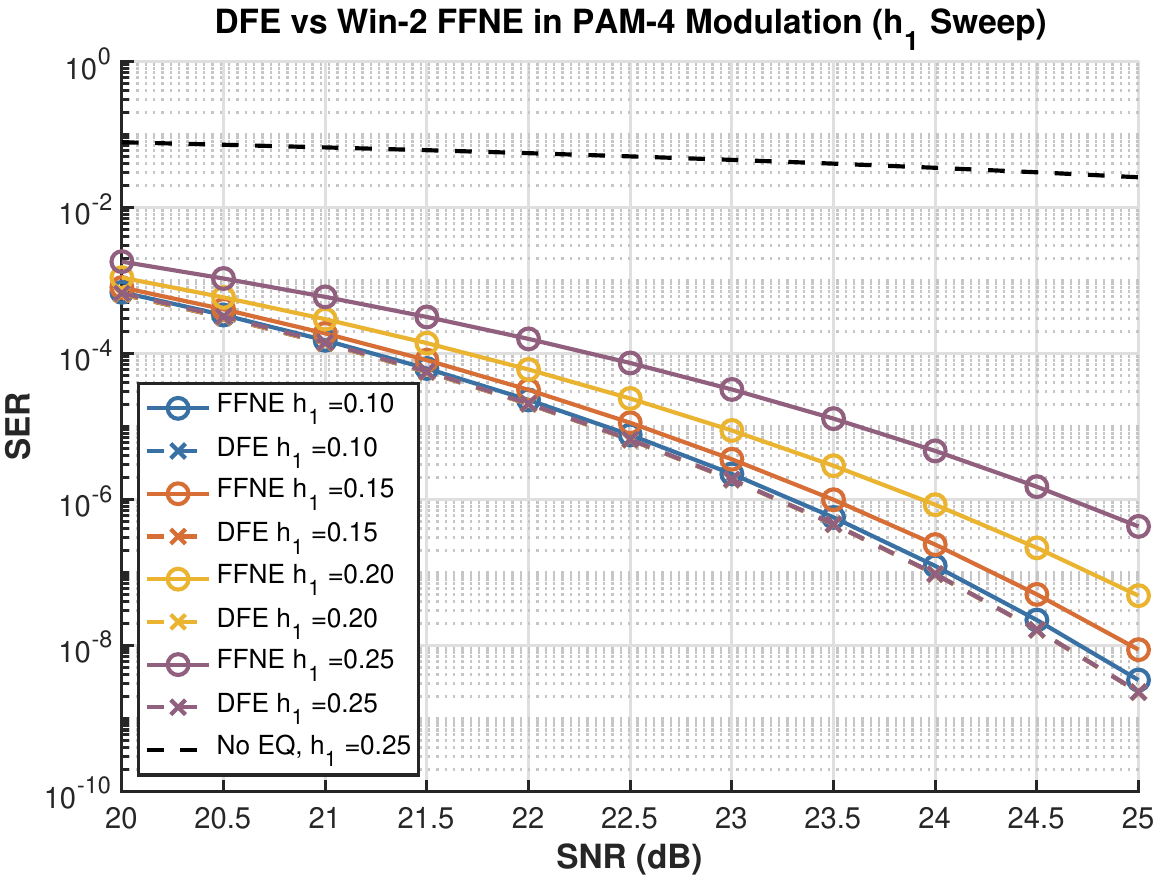}
    \caption{Statistical-model-based SER comparison of the PAM-4 1-tap DFE and the Win-2 FFNE for $h_1$ values from 0.1 to 0.25.}
    \label{fig:dfe_vs_ffmlse_pam4_stat}
\end{figure}

Geometrically, this shifts the sequence-space representation from a 2-D plane to a 3-D Voronoi geometry, as illustrated in Fig. \ref{fig:win3_ffne_decision_cube}. This extra dimension provides greater separation between the critical alternating patterns ([1,0,1,0] and [0,1,0,1]) and the origin, effectively widening the minimum-distance margin and improving noise robustness.

The corresponding hardware implementation is shown in Fig. \ref{fig:win3_diagram}. Relative to the Win-2 FFNE, the Win-3 FFNE requires more than four times as many comparators, leading to substantially higher implementation complexity, area, and power consumption.

As shown in the statistical results of Fig. \ref{fig:dfe_vs_ffne_win3}, the Win-3 FFNE significantly closes the gap with the 1-tap DFE. Unlike the Win-2 version, it maintains near-ideal performance until $h_1$ exceeds approximately 0.4. Even at a high postcursor value of $h_1=0.6$, the SNR penalty remains much smaller than that of the Win-2 FFNE. Detailed technical implementation and source code can be found in the previously cited repository.

As in the Win-2 case, the dominant alternating pattern determines the minimum-distance margin. Its coordinates are
\begin{equation}
\begin{alignedat}{1}
    V_{\mathrm{IN}}[k-2] &= V_{\mathrm{IN}}[k] = +h_0 - h_1, \\
    V_{\mathrm{IN}}[k-1] &= -h_0 + h_1.
\end{alignedat}
\end{equation}
The distance from the origin is
\begin{equation}
\begin{alignedat}{1}
d&=\sqrt{(V_{\mathrm{IN}}[k-2])^2 + (V_{\mathrm{IN}}[k-1])^2 + (V_{\mathrm{IN}}[k])^2}\\
&= \sqrt{3}\lvert h_0-h_1\rvert .
\end{alignedat}
\end{equation}
Equating $d=h_0$ gives $h_1\approx0.423h_0$, extending the Win-2 boundary by $\sqrt{2}$ due to the additional observation dimension. However, this improvement requires substantially higher comparator count, area, and power. Thus, Win-3 mainly serves as an upper-bound example, while Win-2 remains the preferred balance between nonlinear equalization capability and hardware efficiency.

The Win-3 FFNE substantially extends equalization capability, achieving performance comparable to a 1-tap DFE up to $h_1 \approx 0.423$. This represents a $\sqrt{2}$ improvement over the Win-2 version, a gain directly attributed to the increased dimensionality of the 3-D Voronoi decision space.

However, as illustrated in Fig. \ref{fig:win3_diagram}, this performance boost requires a dramatic increase in hardware complexity. While longer window lengths continue to push the equalization boundaries, implementations beyond Win-2 quickly become impractical due to exponential growth in area and power.

Consequently, the Win-3 FFNE serves primarily as a theoretical upper bound, demonstrating the maximum benefit obtainable from expanded windows. For most high-speed serial links, the Win-2 FFNE remains the preferred equalization scheme, providing the best balance between nonlinear equalization performance and hardware efficiency.

\begin{figure*}[b]
\rule{\textwidth}{0.5pt}
\begin{equation}\label{eq:pam4_ffne_stat}
\begin{alignedat}{1}
& P_e=P\left(\hat{a}[k]\neq\frac{1}{3}, \ \big\vert\ a[k]=\frac{1}{3}\right)\\
& \phantom{P_e}=P\left(V_{\mathrm{IN}}[k]<-h_1\right) + P\left(h_1\cdot V_{\mathrm{IN}}[k-1]>V_{\mathrm{IN}}[k], -h_1\leq V_{\mathrm{IN}}[k]<h_1\right) \\
& \phantom{P_e=}+ P\left(h_1\cdot V_{\mathrm{IN}}[k-1]\leq V_{\mathrm{IN}}[k]-V_{TH}, -h_1+V_{TH} \leq V_{\mathrm{IN}}[k]<h_1+V_{TH}\right) + P\left(h_1+V_{TH} \leq V_{\mathrm{IN}}[k] \right)\\ 
& \phantom{P_e}=\Phi\left(\frac{-h_1-\mu_{V_{\mathrm{IN}}[k]}}{\sigma_n}\right)+\int_{-h_1}^{h_1}\int^{\infty}_{\frac{1}{h_1}(V_{\mathrm{IN}}[k])}{\frac{1}{2\pi \sigma_n^2}e^{\left(-\frac{1}{2}\left[S^T\Sigma^{-1}S\right]\right)}}   dV_{\mathrm{IN}}[k-1]dV_{\mathrm{IN}}[k] \\ 
& \phantom{P_e=}+\int_{-h_1+V_{TH}}^{h_1+V_{TH}}\int_{-\infty}^{\frac{1}{h_1}(V_{\mathrm{IN}}[k]-V_{TH})}{\frac{1}{2\pi \sigma_n^2}e^{\left(-\frac{1}{2}\left[S^T\Sigma^{-1}S\right]\right)}}   dV_{\mathrm{IN}}[k-1]dV_{\mathrm{IN}}[k]+Q\left(\frac{h_1+V_{TH}-\mu_{V_{\mathrm{IN}}[k]}}{\sigma_n}\right). \\
\end{alignedat}
\end{equation}
\end{figure*}

\section{Single-tap ($h_1$) Feedforward Nonlinear ISI equalizer for PAM-4 Modulation} \label{sec:sectionIII}
\subsection{Theory of Operation} \label{sec3:subsec1}

In PAM-4 modulation, the transmitted symbols $a[k]$ are drawn from the set $\{-1, -1/3, +1/3, +1\}$. The Win-2 FFNE principle remains identical to the NRZ case: we evaluate all 64 candidate 3-symbol sequences ($4^3$ hypotheses), map them to noiseless points in the $(V_{\mathrm{IN}}[k], V_{\mathrm{IN}}[k-1])$ plane, and select the symbol $\hat{a}[k]$ associated with the minimum-distance hypothesis.

As shown in Fig. \ref{fig:ffmlse_pam4}(a), grouping these results by the final symbol produces four dominant decision bands across the plane. The condition $h_1 \le h_0/3$ characterizes the shallow-loss operating regime targeted by this work; within this regime, the complex 2-D decision boundaries decompose into a set of linear threshold rules:

\begin{align*}
    & \circled{1}:\ V_{\mathrm{IN}}[k]=-h_1-V_{TH} &&\circled{2}:\ V_{\mathrm{IN}}[k]=h_1-V_{TH} \\ 
    & \circled{3}:\ V_{\mathrm{IN}}[k]=-h_1 &&\circled{4}:\ V_{\mathrm{IN}}[k]=h_1 \\ 
    & \circled{5}:\ V_{\mathrm{IN}}[k]=-h_1+V_{TH} &&\circled{6}:\ V_{\mathrm{IN}}[k]=h_1+V_{TH} \\ 
    & \circled{7}:\ V_{\mathrm{IN}}[k-1]=\frac{V_{\mathrm{IN}}[k] + V_{TH}}{h_1} && \circled{8}:\ V_{\mathrm{IN}}[k-1]=\frac{V_{\mathrm{IN}}[k]}{h_1} \\ 
    & \circled{9}:\ V_{\mathrm{IN}}[k-1]=\frac{V_{\mathrm{IN}}[k] - V_{TH} }{h_1}.
\end{align*}
Assuming $h_1\leq h_0\cdot\tfrac{1}{3}$ and $V_{TH}=h_0\cdot \tfrac{2}{3}$, the vertical decision boundaries corresponding to $\circled{2}$ and $\circled{4}$ are always positioned to the left of those corresponding to $\circled{3}$ and $\circled{5}$, as illustrated in Fig. \ref{fig:ffmlse_pam4}(b). Under this condition, both the hardware implementation and the statistical analysis of the Win-2 FFNE for PAM-4 can be further simplified. In addition, the same adaptation procedure used for the NRZ case can be directly applied. 

This formulation allows the PAM-4 Win-2 FFNE to be interpreted as three parallel NRZ Win-2 FFNEs, as illustrated in Fig. \ref{fig:pam4_ffne_diagram}. As a result, the PAM-4 architecture retains the same basic feedforward decision structure as the NRZ case while extending it to four-level signaling. However, as discussed in the next subsection, its noise penalty relative to a DFE increases as the ISI becomes more severe.

\subsection{Statistical Analysis} \label{sec3:subsec2}

\eqref{eq:pam4_ffne_stat} defines the error probability for the case where the transmitter sends the symbol $a[k]=1/3$. As illustrated by the orange region in Fig. \ref{fig:ffmlse_pam4}(a), \eqref{eq:pam4_ffne_stat} evaluates the probability of the received sample falling outside the correct decision boundary. This analysis is extended to the remaining PAM-4 symbols to complete the statistical framework, the full derivation of which is available in the supplementary material. The corresponding simulation source code is hosted at the repository link provided in Section II. Fig. \ref{fig:dfe_vs_ffmlse_pam4_stat} compares the symbol error rate (SER) of the PAM-4 Win-2 FFNE against a 1-tap PAM-4 DFE based on a statistical model. To assess equalizer sensitivity to channel ISI, four postcursor levels are analyzed: $h_1 \in \{0.1, 0.15, 0.2, 0.25\}$.

Two primary trends emerge from Fig. \ref{fig:dfe_vs_ffmlse_pam4_stat}. First, under weak postcursor ISI ($h_1 \leq 0.15$), the Win-2 FFNE performance closely tracks that of the 1-tap DFE across the entire SNR range. This suggests that the 2-D decision regions remain well-separated, with dominant error events comparable to those of an ideal DFE. However, as $h_1$ increases, the FFNE performance begins to deviate, with the gap persisting even at high SNR. Notably, at $h_1 = 0.25$, the FFNE exhibits a significant SNR penalty, consistent with the reduced minimum-distance margin in the sequence-space geometry under heavy ISI. Despite this penalty, the Win-2 PAM-4 FFNE maintains a substantial BER improvement over the unequalized case.

In summary, the Win-2 PAM-4 FFNE trades comparator count for removal of the DFE feedback loop. It avoids direct DFE timing closure, Tx-side pre-shaping, channel-dependent mode switching, and FFE-based $h_1$ cancellation. Its operating envelope is bounded by $h_1\leq h_0/3$; beyond this range, the noise penalty increases and DFE-based or 1+D-shaping approaches become preferable. Longer windows could improve heavy-ISI performance, but their hardware cost scales rapidly with modulation order.

\begin{figure}[t]
    \centering
    \includegraphics[width=0.95\columnwidth]{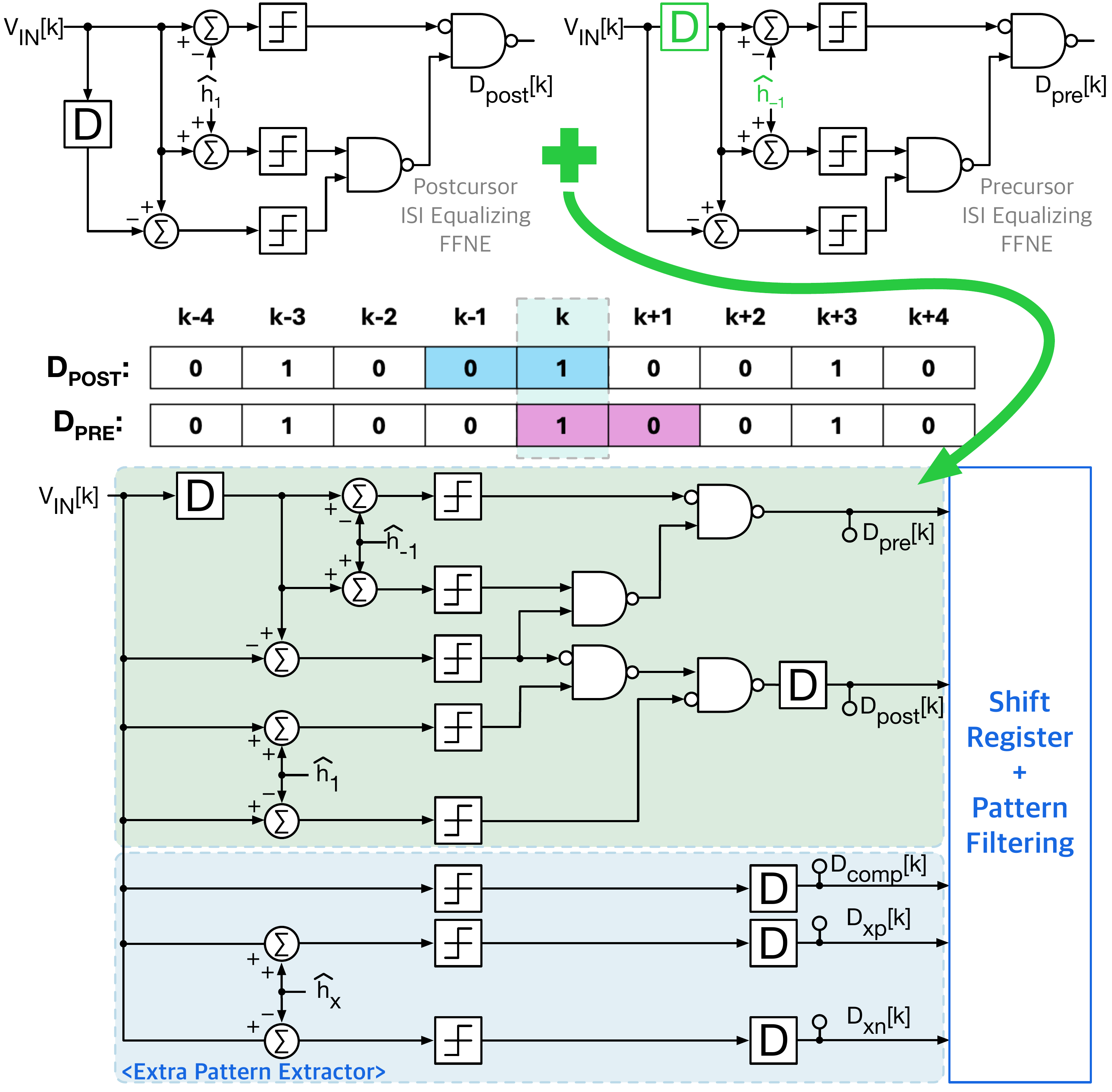}
    \caption{Proposed PD-FFNE architecture. The green-highlighted region implements the Win-2 NRZ FFNE for $h_1$ and $h_{-1}$ equalization. The blue-highlighted region contains additional slicers for extra pattern detection, which substantially improve BER performance. }
    \label{fig:pdffne_prepost}
\end{figure}

\section{Double-tap ($h_{-1}\ \&\ h_{1}$) Feedforward Nonlinear ISI equalizer for NRZ Modulation} \label{sec:sectionIV}

Although multi-tap FFNEs can in principle be derived by extending the Voronoi-diagram-based formulation, their practical value quickly diminishes because hardware complexity grows rapidly with tap count, similar to the scaling observed for higher-order modulation in the PAM-4 Win-2 FFNE. A more practical alternative is obtained by combining the precursor-ISI and postcursor-ISI Win-2 FFNE structures in a symmetric feedforward architecture and augmenting them with pattern detection and filtering. This leads to the proposed pattern-detection-based FFNE (PD-FFNE), which improves equalization capability while avoiding any feedback-loop timing constraint.

It is worth noting that prior work in \cite{yusang_dicode} also proposed a feedforward equalization technique. However, that approach relies on transmitter-side encoding, making it unsuitable for backward-compatible deployment in most SerDes links, and its performance remains inferior to that of a conventional DFE. The extended version in \cite{osu_ml} demonstrated improved performance, but it likewise requires transmitter-side encoding and does not report a comparison against DFE. In contrast, the proposed PD-FFNE requires no transmitter-side encoding while outperforming a conventional FFE+DFE baseline.

This section presents the operating principle, circuit implementation, adaptation method, and performance of the proposed PD-FFNE.

\begin{figure}[t]
    \centering
    \captionsetup[subfigure]{justification=centering}
        \begin{subfigure}[b]{\columnwidth}
        \centering
        \includegraphics[width=0.85\columnwidth]{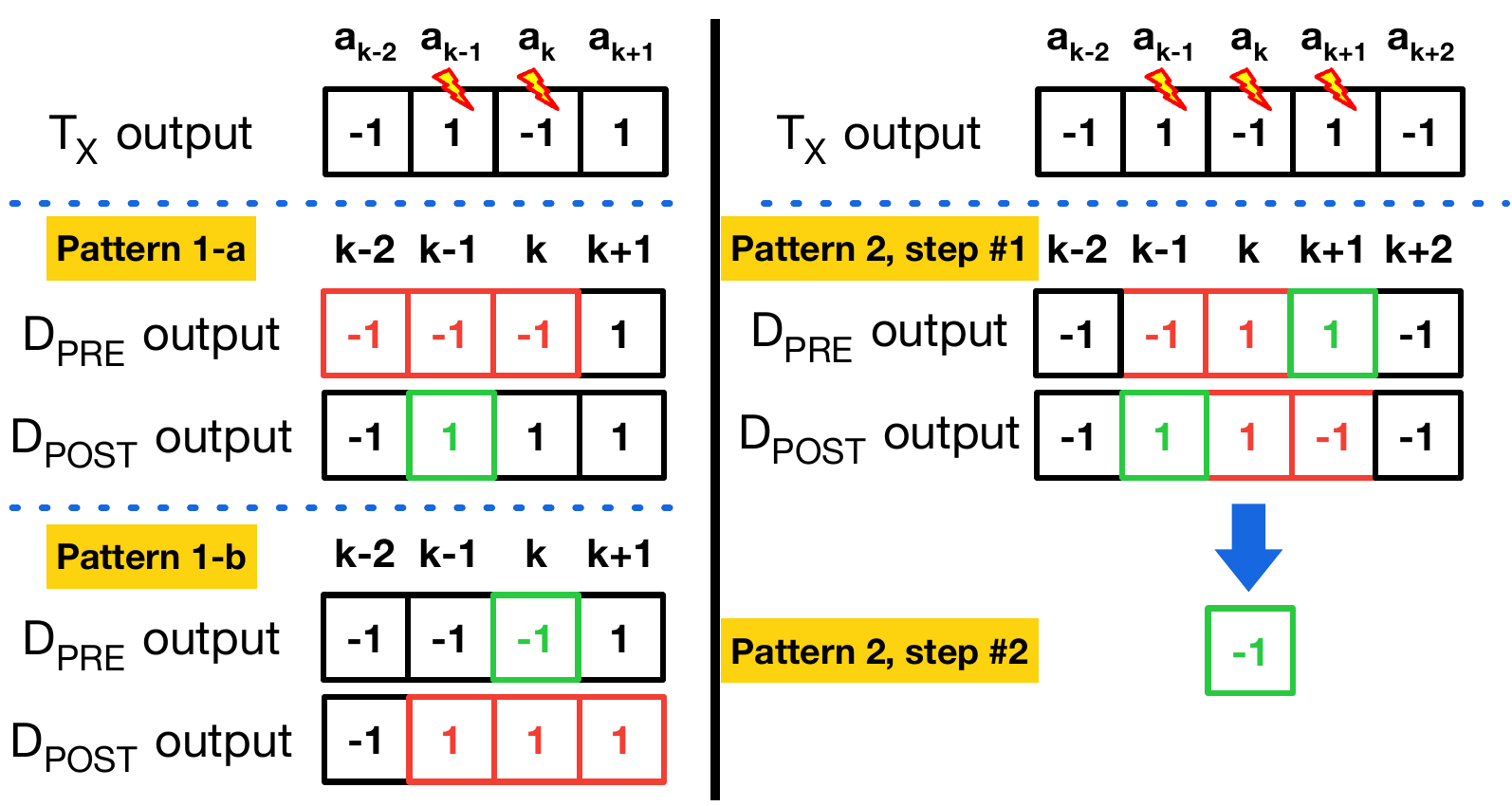}
        \caption{}
        \end{subfigure}
        
        \hfill
        \begin{subfigure}[b]{\columnwidth}
        \centering
        \includegraphics[width=0.85\columnwidth]{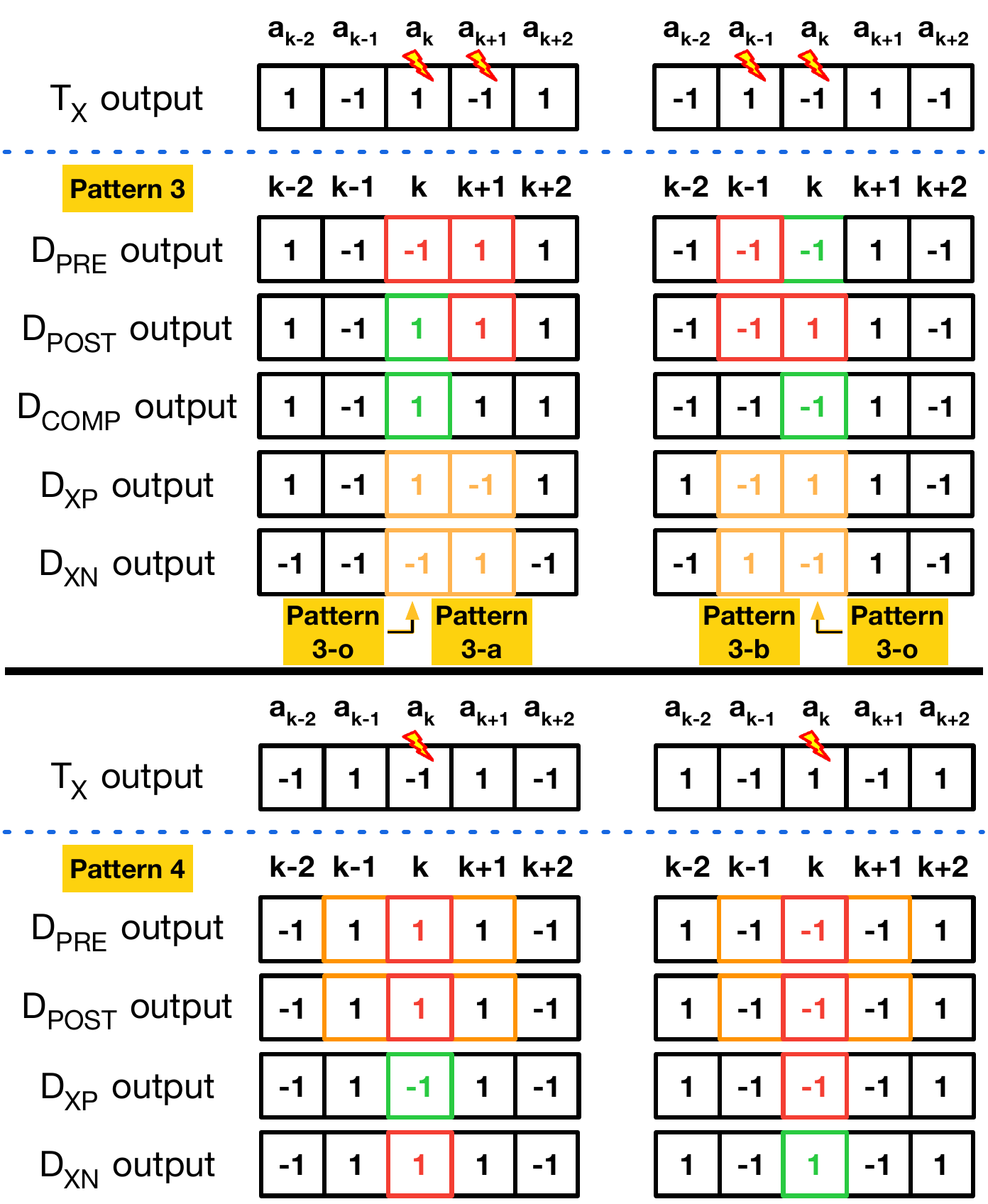}
        \caption{}
        \end{subfigure}

    \caption{(a) Patterns 1 and 2 realized by the $h_{-1}$- and $h_1$-equalizing window-length-2 FFNE. (b) Pattern 3 and 4 detection via the extra pattern extractor.}
    \label{fig:pdffne_patterns}
\end{figure}

\begin{figure}[!t]
    \centering
    \captionsetup[subfigure]{justification=centering}
        \begin{subfigure}[b]{\columnwidth}
        \centering
        \includegraphics[width=0.85\columnwidth]{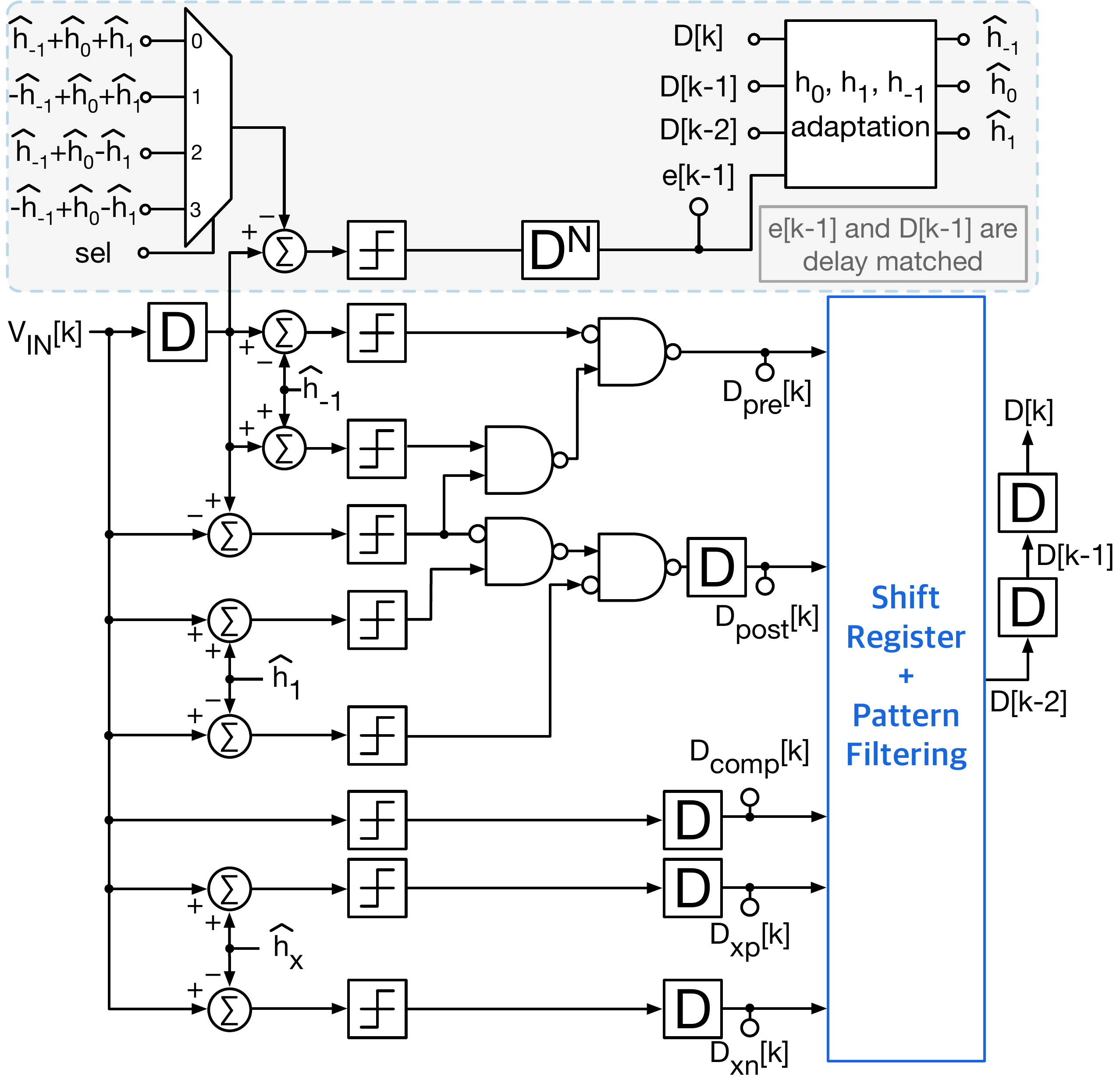}
        \caption{}
        \end{subfigure}
        
        \begin{subfigure}[b]{\columnwidth}
        \centering
        \includegraphics[width=0.85\columnwidth]{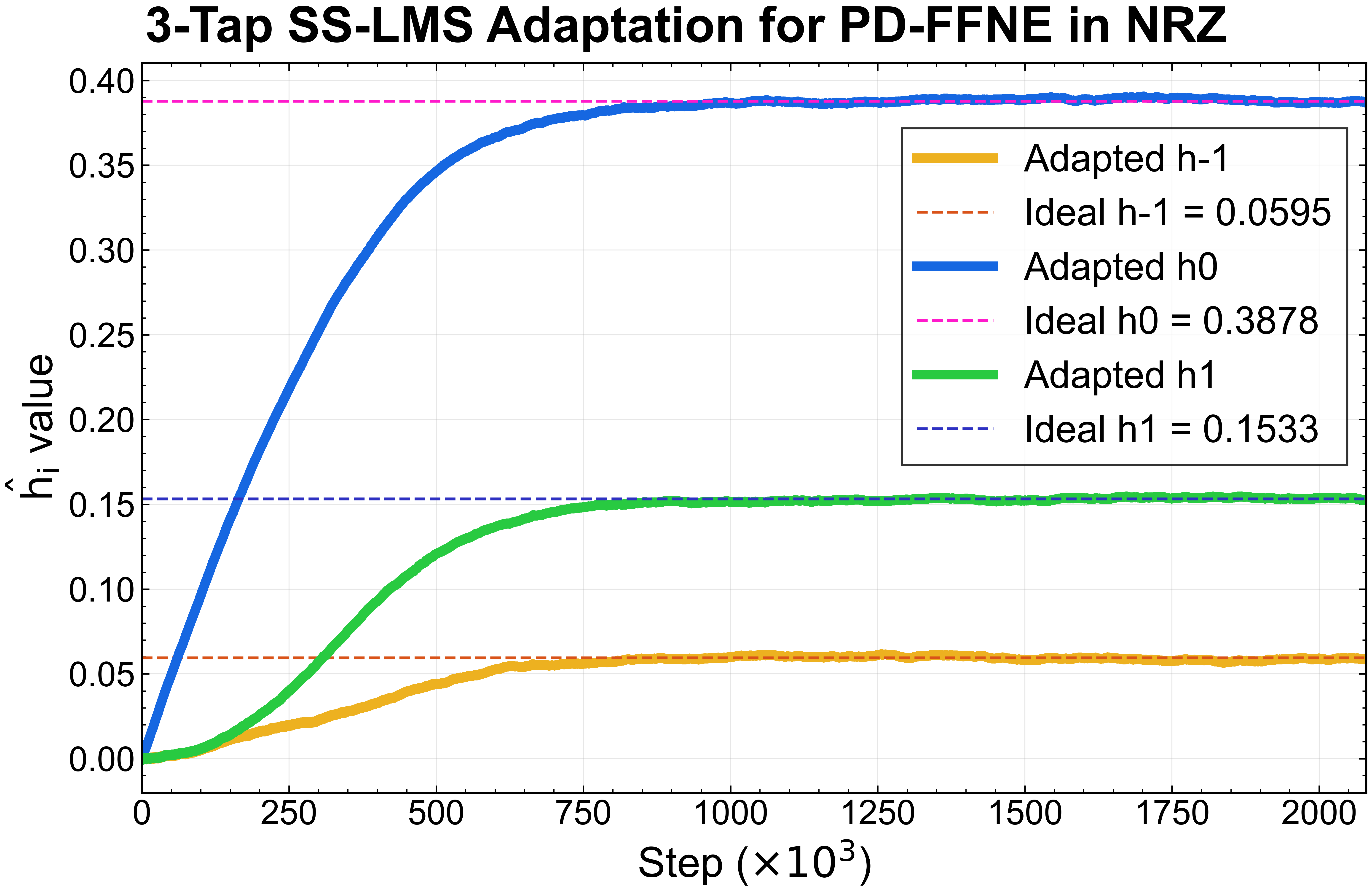}
        \caption{}
        \end{subfigure}

    \caption{(a) Error-slicer implementation for $\hat{h}_{-1}$ and $\hat{h}_{1}$ adaptation in PD-FFNE. (b) SS-LMS convergence of the adapted tap coefficients.}
    \label{fig:pdmlse_adaptation}
\end{figure}

\subsection{Theory of Operation} \label{sec4:subsec1}

The FFNE structure developed in Section \ref{sec:sectionII} can also be rearranged to cancel precursor ISI ($h_{-1}$) by relocating the delay element and reassigning the tap coefficients, as illustrated in the upper-right portion of Fig. \ref{fig:pdffne_prepost}. Once this precursor-equalizing variant is recognized, a natural extension is to combine it with the original postcursor-equalizing FFNE so that both $h_{-1}$ and $h_1$ can be equalized simultaneously. A straightforward implementation instantiates the two substructures in parallel and aligns their outputs to the same symbol index $k$, leading to the combined architecture shown in the green-highlighted region of Fig. \ref{fig:pdffne_prepost}.

Depending on the channel ISI profile and input noise level, the outputs \dpre and \dpost do not always agree. Through heuristic analysis of these decision-conflict events, two useful correction patterns were identified using the two FFNE branches alone, and two additional patterns were enabled by the extra detectors shown in the blue-highlighted region of Fig. \ref{fig:pdffne_prepost}. This PD-FFNE consists of a symmetric dual-FFNE core augmented by a lightweight pattern-extraction and filtering block. Fig. \ref{fig:pdffne_patterns} summarizes the detected conflict signatures.

Consider pattern 1-a, where the received voltage samples at $k-1$ and $k$ are contaminated by noise. The first decision policy is triggered if D\textsubscript{PRE} (or D\textsubscript{POST}) has three consecutive identical bits (from $k-2$ to $k$ in this case), and \dpre and \dpost have a bit conflict at $k-1$. If decisions made by \dpre from $k-2$ to $k$ are correct, then the magnitude of the voltage sample at $k-1$ is on the order of $h_{-1}+h_0+h_1$. Because this value is large, a disagreement at $k-1$ should be unlikely unless an unusual strong noise perturbation is present. Therefore, observing a conflict at $k-1$ suggests that the branch producing the consistent three-bit pattern (\dpre decision) is more likely wrong. Pattern 1-b follows the same reasoning as pattern 1-a, but corresponds to the case in which \(\dpost\) is incorrect.

Pattern 2 is triggered when the \dpre and \dpost disagree at $k-1$ and $k+1$, but produce the same decision at $k$. Such a pattern arises when the received samples from $k-1$ to $k+1$ are contaminated by noise, with the sample at $k$ severely contaminated. Starting with the bit conflict at $k+1$, \dpost is unreliable since its decision for $k+1$ strongly depends on the contaminated sample at $k$. Thus, \dpre is treated as the more reliable for the symbol at $k+1$. By the same reason, \dpost is a more reliable decision for the $k-1$ sample. Lastly, since these conflicts are mainly caused by the contaminated sample at $k$, we simply negate the decision for $k$ even if \dpre and \dpost outputs at $k$ have no conflict.  

To further improve the correction capability, three additional detectors are introduced, as shown in the blue-highlighted region of Fig. \ref{fig:pdffne_prepost}. These detectors allow two additional conflict patterns to be identified. The auxiliary coefficient $h_x$ is treated as a fixed detector parameter and is nominally chosen as $h_0/2$. This choice provides a simple channel-independent setting for the proposed PD-FFNE without requiring an additional adaptation loop. A sensitivity study using the target channel response is presented in the next section, where the impact of $h_x$ on the BER performance is quantified.

\begin{figure}[t]
    \centering
    \includegraphics[width=0.95\columnwidth]{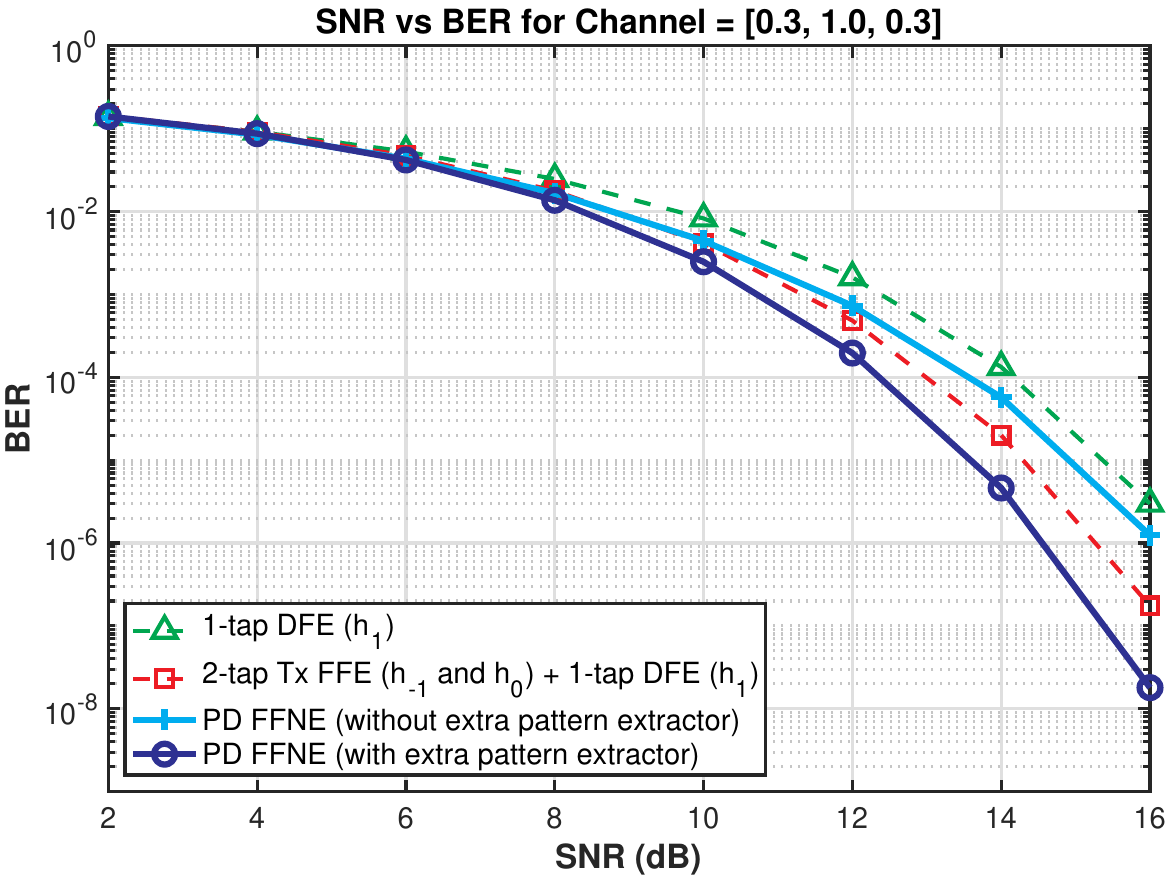}
    \caption{BER versus SNR for channel [0.3,\,1.0,\,0.3], demonstrating that the proposed improved PD-FFNE with extra pattern detection outperforms conventional practical baselines, particularly FFE+DFE, under controlled ISI conditions.}
    \label{fig:pdmlse_ber_comparison}
\end{figure}

Pattern 3 is triggered when \dpre and \dpost disagree at $k$ and agree at both $k+1$ and $k-1$. In addition, \dxp and \dxn disagree at $k$ and at either $k-1$ or $k+1$. This condition occurs when the sample at $k+1$ (or $k-1$) is strongly corrupted by noise, while the sample at $k$ experiences modest perturbation. The first signature is a decision mismatch between \dpre and \dpost at index $k$. We then use one of the added detectors, \dcomp, as a tie-breaker: among \dpre and \dpost, we select the decision that matches \dcomp. In the example shown in the upper-right side of Fig. \ref{fig:pdffne_patterns}, \dcomp\ agrees with \dpre; hence, \dpre is taken as the more reliable decision at $k$. This outcome implies that the dominant impairment is the severely corrupted sample at $k-1$, which propagates into the postcursor-based path and causes \dpost to fail at $k$. Given this diagnosis, we further evaluate the reliability of the decision at $k-1$, even if \dpre and \dpost do not explicitly disagree at that index. This is where \dxp and \dxn are used. If \dxp = -1 and \dxn = 1 at $k-1$, it indicates a large noise excursion on the $k-1$ sample. In this case, we flip the decisions of \dpre and \dpost at $k-1$.

\begin{figure*}[!t]
    \centering
    \includegraphics[width=\linewidth]{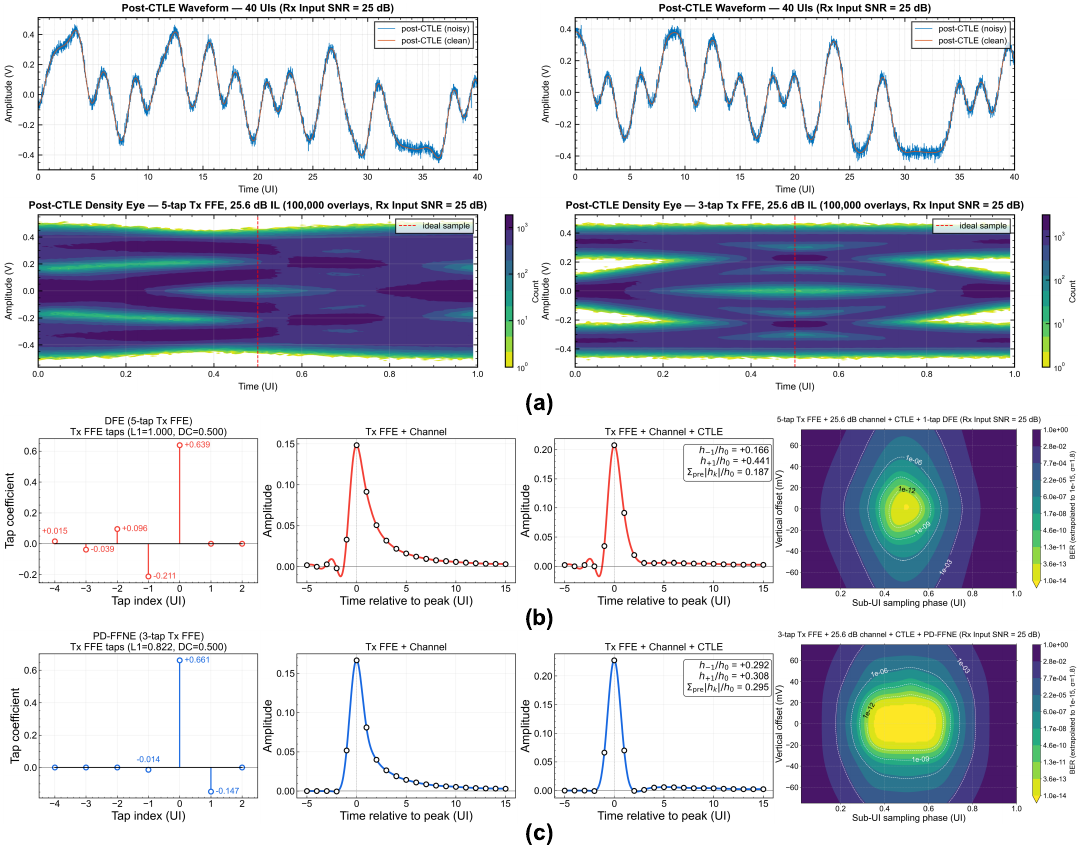}
    \caption{Detector input waveform, density eye, Tx-FFE taps, pulse responses, cursor ratios, and BER contour for the 25.6-dB-loss channel at 25-dB Rx input-referred SNR. (a) Detector input waveform and density eye. (b) 5-tap Tx-FFE + 1-tap DFE baseline. (c) 3-tap Tx-FFE + PD-FFNE. With the same Tx-FFE DC gain of 0.5, PD-FFNE widens the BER margin while reducing $\lVert\vec{c}_{\mathrm{TX}}\rVert_1$ from 1.000 to 0.822.}
    \label{fig:pdmlse_hvmargin}
\end{figure*}

\begin{figure}[t]
    \centering
    \includegraphics[width=0.9\columnwidth]{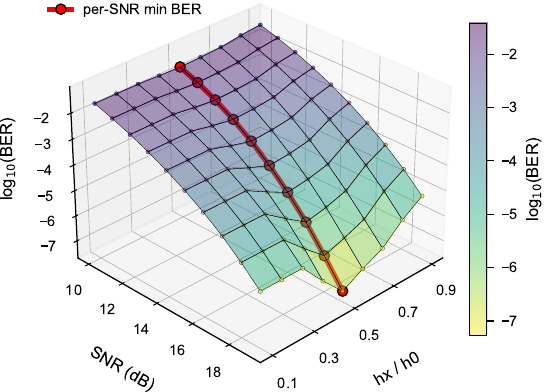}
    \caption{$\log_{10}(\mathrm{BER})$ versus SNR and $h_x/h_0$ for the Tx-FFE + PD-FFNE receiver on the 25.6-dB-loss channel.}
    \label{fig:pdmlse_hx_sweep}
\end{figure}

\begin{figure}[t]
    \centering
    \includegraphics[width=0.9\columnwidth]{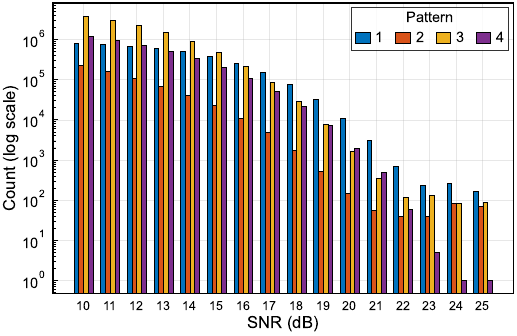}
    \caption{PD-FFNE pattern-firing histogram versus SNR for the 25.6-dB channel with a 3-tap Tx-FFE over $10^8$ symbols per SNR point.}
    \label{fig:pdmlse_pattern_histogram}
\end{figure}

Pattern 4 is triggered when \dpre and \dpost produce identical decisions over the three-symbol window $(k-1,k,k+1)$, while \dxp and \dxn disagree at $k$. This condition indicates that the voltage sample at $k$ is likely corrupted. The intuition mirrors Pattern 1. If \dpre and \dpost are correct and exhibit three consecutive identical decisions, the noiseless sample magnitude at $k$ should be $|h_{-1}+h_0+h_1|$. Such a large-amplitude sample is robust to noise. Therefore, observing a conflict between \dxp and \dxn suggests that the sample at $k$ has been significantly perturbed and may have simultaneously flipped the decisions of both \dpre and \dpost\!\!.

The tap coefficients of the PD-FFNE can be adapted using a $dLev$-based approach similar to that used for the Win-2 FFNE in Section \ref{sec:sectionII}. \eqref{eq:pdffne_adaptation1} and \eqref{eq:pdffne_adaptation2} define the corresponding updates, and Fig. \ref{fig:pdmlse_adaptation} shows both the error-slicer implementation and the convergence behavior of the adapted tap coefficients for the same channel considered in Fig. \ref{fig:win2mlse_adaptation_plot}.
\begin{equation}\label{eq:pdffne_adaptation1}\
\begin{alignedat}{1}
    dLev_{111}^{k} &= dLev_{111}^{k-1} + \mu\, e[k-1]\, D[k-1] \\ 
    &\text{ update only when }\{D[k-2:k]\}\myeq\{1,1,1\},\\ 
    dLev_{010}^{k} &= dLev_{010}^{k-1} + \mu\, e[k-1]\, D[k-1] \\ 
    & \text{ update only when }\{D[k-2:k]\}\myeq\{0,1,0\},\\ 
    dLev_{110}^{k} &= dLev_{110}^{k-1} + \mu\, e[k-1]\, D[k-1] \\ 
    & \text{ update only when }\{D[k-2:k]\}\myeq\{1,1,0\},\\ 
    dLev_{011}^{k} &= dLev_{011}^{k-1} + \mu\, e[k-1]\, D[k-1] \\ 
    & \text{ update only when }\{D[k-2:k]\}\myeq\{0,1,1\}.\\ 
\end{alignedat}
\end{equation}
\begin{equation}\label{eq:pdffne_adaptation2}\
\begin{alignedat}{1}
    \hat{h}_0^k &= \frac{dLev_{111}^k + dLev_{010}^k + dLev_{110}^k + dLev_{011}^k}{4} ,\\
    \hat{h}_1^k &= \frac{dLev_{111}^k - dLev_{010}^k + \left(dLev_{011}^k - dLev_{110}^k\right)}{4}, \\ 
    \hat{h}_{-1}^k &= \frac{dLev_{111}^k - dLev_{010}^k - \left(dLev_{011}^k - dLev_{110}^k\right)}{4} .
\end{alignedat}
\end{equation}

\subsection{Performance Comparison} \label{sec4:subsec2}

This section evaluates the proposed PD-FFNE in terms of BER, eye-margin robustness, Tx-FFE coefficient utilization, and sensitivity to the auxiliary detector coefficient. The results show that PD-FFNE improves error-rate performance and timing robustness over DFE-based baselines while maintaining low hardware complexity.

Fig. \ref{fig:pdmlse_ber_comparison} compares the BER performance for the artificial channel $[h_{-1},h_0,h_1]=[0.3,1.0,0.3]$. The improved PD-FFNE with the additional pattern detectors, highlighted in blue, achieves the lowest BER over the full simulated SNR range. For SNR values above 15 dB, it provides approximately an order-of-magnitude BER reduction relative to the conventional FFE+DFE baseline, corresponding to at least 1-dB SNR gain.

Fig. \ref{fig:pdmlse_hvmargin} compares the proposed PD-FFNE configuration with a conventional 5-tap Tx-FFE + 1-tap DFE baseline for a 25.6-dB-loss channel at the Nyquist frequency. A simplified CTLE with a zero at 8 GHz and poles at 20 GHz and 50 GHz is included at the receiver, while no Rx-FFE is used so that the comparison remains focused on Rx-side nonlinear postcursor correction without introducing additional adaptation dimensions. The Rx input-referred SNR is intentionally set to 25 dB to create a partially noise-closed eye, allowing the BER contours to capture the combined impact of residual ISI and additive noise. Both configurations use the same Tx-FFE DC gain of 0.5, enforcing the same nominal receiver input swing. The baseline Tx-FFE is optimized to minimize precursor ISI subject to $\lVert\vec{c}_{\mathrm{TX}}\rVert_1\leq 1$, while the 1-tap DFE cancels the dominant first postcursor. In contrast, the proposed configuration uses a 3-tap Tx-FFE to balance precursor and postcursor ISI while preserving the residual ISI structure required by PD-FFNE.

The resulting Tx-FFE coefficients, pulse responses, and BER contours show two key advantages of PD-FFNE. First, the rightmost BER contours exhibit a wider low-BER region in both sampling-phase and vertical-offset dimensions, corresponding to improved horizontal and vertical eye margins. Second, PD-FFNE reduces Tx-FFE coefficient utilization: the baseline fully consumes the available Tx-FFE coefficient budget, $\lVert\vec{c}_{\mathrm{TX}}\rVert_1=1$, whereas the proposed configuration preserves coefficient headroom under the same DC-gain constraint. This headroom arises because PD-FFNE only requires the Tx-FFE to reduce $h_{-1}/h_0$ below the limit derived in \eqref{eq:win2_h1_limit}, corresponding to $h_{-1}/h_0<0.3$ in this example, rather than aggressively zero-forcing $h_{-1}$. Consequently, the remaining coefficient budget can be used for actual channel ISI instead of compensating secondary precursor components at $h_{-2}$--$h_{-4}$ introduced by aggressive precursor suppression.

Fig. \ref{fig:pdmlse_hx_sweep} evaluates the sensitivity of PD-FFNE to the auxiliary coefficient $h_x$ used in \dcomp. Using the same channel and equalization configuration as in Fig. \ref{fig:pdmlse_hvmargin}, the Rx input-referred SNR is swept from 10 dB to 19 dB, while $h_x/h_0$ is swept from 0.1 to 0.9. The BER surface exhibits a clear minimum near $h_x/h_0=0.5$ across the simulated SNR range, indicating that the auxiliary detector is most effective when its threshold is placed near the midpoint of the main-cursor amplitude. Since the per-SNR minimum remains close to this value, precise adaptation of $h_x$ is unnecessary; the fixed choice $h_x=h_0/2$ therefore provides a robust nominal setting.

Fig. \ref{fig:pdmlse_pattern_histogram} shows the PD-FFNE pattern-firing histogram under the same channel and equalization condition as in Fig. \ref{fig:pdmlse_hvmargin}, with SNR swept over the specified range. At low SNR values, Pattern 3 is the most frequently triggered correction event. As SNR increases, Pattern 1 becomes dominant, while Patterns 2 and 3 occur with comparable frequency at high SNR. Pattern 4 occurs less frequently as SNR improves, but it is retained because rare correction events can still provide measurable BER benefit when they correspond to otherwise uncorrected error mechanisms.

The detailed hardware implementation methodology is provided in the supplementary material, and example SystemVerilog RTL is available in the accompanying repository\footnote{\url{https://github.com/kunmok/digital_backend_for_128GSPS_ADC/tree/main/pdmlse_top/rtl/dsp_backend}.}. The Python and MATLAB code for simulation are also available in the previous footnote.

\section{Conclusion} 
\label{sec:conclusion}

This paper introduced a feedforward nonlinear equalizer (FFNE) framework that removes the feedback-loop timing bottleneck of DFE-based equalization while preserving the noise advantage of nonlinear ISI cancellation. The Win-2 FFNE was derived as a hardware-simplified, fully feedforward form of short-window maximum-likelihood sequence estimation, with practical implementation, adaptation, and statistical analysis. The resulting framework revealed a fundamental Win-2 limitation: the dominant alternating sequence points collapse toward the origin, producing an SNR penalty relative to an ideal 1-tap DFE beyond $h_1 \approx 0.293h_0$.

FFNE was further clarified as a performance-to-complexity design space. Win-3 improves sequence-point separation at sharply increased hardware cost, while the multi-level extension confirms that the same geometric framework remains applicable beyond binary signaling. The PD-FFNE further improves ISI mitigation by combining precursor- and postcursor-oriented Win-2 FFNE structures with pattern detection and filtering. Under representative channel conditions, it outperforms the conventional FFE+DFE baseline while preserving a fully feedforward structure and avoiding transmitter-side precoding.

Overall, FFNE provides a practical nonlinear equalization framework for regimes where feedback timing closure, interleaving scalability, and compact analog-and-mixed-signal realization are key constraints. PD-FFNE strengthens this framework with additional feedforward comparators and pattern-detection logic, improving ISI mitigation without reintroducing feedback timing loops.

\bibliographystyle{IEEEtran}
\bibliography{./IEEEabrv,./references}

@STRING{IEEE_J_JSSC       = "{IEEE} J. Solid-State Circuits"}

@article{Bailey2022,
author = {Bailey, James and Shakiba, Hossein and Nir, Ehud and Marderfeld, Grigory and Krotnev, Peter and Lacroix, Marc Andre and Cassan, David and Tonietto, Davide},
journal = IEEE_J_JSSC,
title = {{A 112-Gb/s PAM-4 Low-Power Nine-Tap Sliding-Block DFE in a 7-nm FinFET Wireline Receiver}},
number = {1},
pages = {32--43},
volume = {57},
year = {2022}
}

@ARTICLE{kunmo_sbdfe,
  author={Kim, Kunmo and Moon, Suhong and Han, Jaeduk and Alon, Elad and Niknejad, Ali M.},
  journal={IEEE Transactions on Circuits and Systems I: Regular Papers}, 
  title={{Precursor ISI Cancellation Sliding-Block DFE for High-Speed Wireline Receivers}}, 
  year={2023},
  volume={70},
  number={10},
  pages={4169-4182},
  doi={10.1109/TCSI.2023.3298954}
}

@ARTICLE{emami_ffne,
  author={Emami Meybodi, Mohammad and Shakiba, Hossein and Sheikholeslami, Ali},
  journal={IEEE Open Journal of Circuits and Systems}, 
  title={{Analysis and Design of an Optimal Noise Estimation and Cancellation Filter in Wireline Communication}}, 
  year={2024},
  volume={5},
  number={},
  pages={153-165},
  doi={10.1109/OJCAS.2024.3391698}
}

@phdthesis{odling_thesis,
    Author= {\"Odling, Per},
    Title= {{Design and Analysis of Digital Receivers}},
    School= {{Division of Signal Processing, Luleå University of Technology, Luleå, Sweden}},
    Year= {1995},
    Month= {May},
    Url= {https://www.diva-portal.org/smash/get/diva2:999751/FULLTEXT01.pdf}, 
    Number= {1995:165 D}
}

@ARTICLE{vlad_adaptation,
  author={Stojanovic, V. and Ho, A. and Garlepp, B.W. and Chen, F. and Wei, J. and Tsang, G. and Alon, E. and Kollipara, R.T. and Werner, C.W. and Zerbe, J.L. and Horowitz, M.A.},
  journal=IEEE_J_JSSC, 
  title={{Autonomous dual-mode (PAM2/4) serial link transceiver with adaptive equalization and data recovery}}, 
  year={2005},
  volume={40},
  number={4},
  pages={1012-1026},
  doi={10.1109/JSSC.2004.842863}
}

@ARTICLE{lsi_25gbps,
  author={Kimura, Hiroshi and Aziz, Pervez M. and Jing, Tai and Sinha, Ashutosh and Kotagiri, Shiva Prasad and Narayan, Ram and Gao, Hairong and Jing, Ping and Hom, Gary and Liang, Anshi and Zhang, Eric and Kadkol, Aniket and Kothari, Ruchi and Chan, Gordon and Sun, Yehui and Ge, Benjamin and Zeng, Jason and Ling, Kathy and Wang, Michael C. and Malipatil, Amaresh and Li, Lijun and Abel, Christopher and Zhong, Freeman},
  journal=IEEE_J_JSSC, 
  title={{A 28 Gb/s 560 mW Multi-Standard SerDes With Single-Stage Analog Front-End and 14-Tap Decision Feedback Equalizer in 28 nm CMOS}}, 
  year={2014},
  volume={49},
  number={12},
  pages={3091-3103},
  doi={10.1109/JSSC.2014.2349974}}

@article{ihp_dfe,
  title={Design and measurement techniques for an 80 Gb/s 1-tap decision feedback equalizer},
  author={Awny, Ahmed and Moeller, Lothar and Junio, Joseph and Scheytt, J Christoph and Thiede, Andreas},
  journal={IEEE Journal of Solid-State Circuits},
  volume={49},
  number={2},
  pages={452--470},
  year={2013},
  publisher={IEEE}
}

@ARTICLE{dfe_loopunroll_ibm,
  author={Ozkaya, Ilter and Cevrero, Alessandro and Francese, Pier Andrea and Menolfi, Christian and Morf, Thomas and Brändli, Matthias and Kuchta, Daniel M. and Kull, Lukas and Baks, Christian W. and Proesel, Jonathan E. and Kossel, Marcel and Luu, Danny and Lee, Benjamin G. and Doany, Fuad E. and Meghelli, Mounir and Leblebici, Yusuf and Toifl, Thomas},
  journal={IEEE Journal of Solid-State Circuits}, 
  title={A 64-Gb/s 1.4-pJ/b NRZ Optical Receiver Data-Path in 14-nm CMOS FinFET}, 
  year={2017},
  volume={52},
  number={12},
  pages={3458-3473},
  doi={10.1109/JSSC.2017.2734913}}

@ARTICLE{dfe_loopunroll,
  author={Kasturia, S. and Winters, J.H.},
  journal={IEEE Journal on Selected Areas in Communications}, 
  title={{Techniques for High-Speed Implementation of Nonlinear Cancellation}}, 
  year={1991},
  volume={9},
  number={5},
  pages={711-717},
  doi={10.1109/49.87640}}

@INPROCEEDINGS{dffe,
  author={Pola, Ariel L. and Crivelli, Diego E. and Cousseau, Juan E. and Agazzi, Oscar E. and Hueda, Mario R.},
  booktitle={2011 IEEE International Symposium of Circuits and Systems (ISCAS)}, 
  title={{A new low complexity iterative equalization architecture for high-speed receivers on highly dispersive channels: Decision feedforward equalizer (DFFE)}}, 
  year={2011},
  volume={},
  number={},
  pages={133-136},
  doi={10.1109/ISCAS.2011.5937519}}

@ARTICLE{xilinx_dfe,
  author={Im, Jay and Freitas, Dave and Roldan, Arianne Bantug and Casey, Ronan and Chen, Stanley and Chou, Chuen-Huei Adam and Cronin, Tim and Geary, Kevin and McLeod, Scott and Zhou, Lei and Zhuang, Ian and Han, Jaeduk and Lin, Sen and Upadhyaya, Parag and Zhang, Geoff and Frans, Yohan and Chang, Ken},
  journal=IEEE_J_JSSC, 
  title={{A 40-to-56 Gb/s PAM-4 Receiver With Ten-Tap Direct Decision-Feedback Equalization in 16-nm FinFET}}, 
  year={2017},
  volume={52},
  number={12},
  pages={3486-3502},
  doi={10.1109/JSSC.2017.2749432}}

@ARTICLE{broadcom_dfe,
  author={Zhang, Bo and Vasani, Anand and Sinha, Ashutosh and Nilchi, Alireza and Tong, Haitao and Rao, Lakshmi P. and Khanoyan, Karapet and Hatamkhani, Hamid and Yang, Xiaochen and Meng, Xin and Wong, Alexander and Kim, Jun and Jing, Ping and Sun, Yehui and Nazemi, Ali and Liu, Dean and Brewster, Anthony and Cao, Jun and Momtaz, Afshin},
  journal=IEEE_J_JSSC, 
  title={{A 112-Gb/s Serial Link Transceiver With Three-Tap FFE and 18-Tap DFE Receiver for up to 43-dB Insertion Loss Channel in 7-nm FinFET Technology}}, 
  year={2024},
  volume={59},
  number={1},
  pages={8-18},
  doi={10.1109/JSSC.2023.3313524}}

@ARTICLE{yusang_dicode,
  author={Chun, Yusang and Anand, Tejasvi},
  journal={IEEE Journal of Solid-State Circuits}, 
  title={{An ISI-Resilient Data Encoding for Equalizer-Free Wireline Communication—Dicode Encoding and Error Correction for 24.2-dB Loss With 2.56 pJ/bit}}, 
  year={2020},
  volume={55},
  number={3},
  pages={567-579},
  doi={10.1109/JSSC.2019.2959487}}

@INPROCEEDINGS{javadi_nrz,
  author={Javadi, Ramin and Lin, Xiaohui and Anand, Tejasvi},
  booktitle={2025 Symposium on VLSI Technology and Circuits (VLSI Technology and Circuits)}, 
  title={A 3.2pJ/b 0.068pJ/b/dB 25Gb/s NRZ Wireline Transceiver with 3-Tap FFE and Random Forest Classification for Compensating 47dB Loss in 16nm FinFET}, 
  year={2025},
  volume={},
  number={},
  pages={1-3},
  doi={10.23919/VLSITechnologyandCir65189.2025.11075116}}

@inproceedings{turker_dfe,
  title={A 19Gb/s 38mW 1-tap speculative DFE receiver in 90nm CMOS},
  author={Turker, Didem Z and Rylyakov, Alexander and Friedman, Daniel and Gowda, Sudhir and Sanchez-Sinencio, Edgar},
  booktitle={2009 Symposium on VLSI Circuits},
  pages={216--217},
  year={2009},
  organization={IEEE}
}

@article{vlad_dfe,
  title={Autonomous dual-mode (PAM2/4) serial link transceiver with adaptive equalization and data recovery},
  author={Stojanovic, Vladimir and Ho, Andrew and Garlepp, Bruno W and Chen, Fred and Wei, Jason and Tsang, Grace and Alon, Elad and Kollipara, Ravi T and Werner, Carl W and Zerbe, Jared L and others},
  journal={IEEE Journal of Solid-State Circuits},
  volume={40},
  number={4},
  pages={1012--1026},
  year={2005},
  publisher={IEEE}
}

@INPROCEEDINGS{osu_ml,
  author={Wang, Zhiping and Megahed, Mohamed and Chun, Yusang and Anand, Tejasvi},
  booktitle={2021 Symposium on VLSI Circuits}, 
  title={{A Machine Learning Inspired Transceiver with ISI-Resilient Data Encoding: Hybrid-Ternary Coding + 2-Tap FFE + CTLE + Feature Extraction and Classification for 44.7dB Channel Loss in 7.3pJ/bit}}, 
  year={2021},
  volume={},
  number={},
  pages={1-2},
  doi={10.23919/VLSICircuits52068.2021.9492510}}

@INPROCEEDINGS{ibm_100gbps,
  author={Cevrero, Alessandro and Ozkaya, Ilter and Francese, Pier Andrea and Brandli, Matthias and Menolfi, Christian and Morf, Thomas and Kossel, Marcel and Kull, Lukas and Luu, Danny and Dazzi, Martino and Toifl, Thomas},
  booktitle={2019 IEEE International Solid-State Circuits Conference - (ISSCC)}, 
  title={{6.1 A 100Gb/s 1.1pJ/b PAM-4 RX with Dual-Mode 1-Tap PAM-4 / 3-Tap NRZ Speculative DFE in 14nm CMOS FinFET}}, 
  year={2019},
  volume={},
  number={},
  pages={112-114},
  doi={10.1109/ISSCC.2019.8662495}}

@ARTICLE{aurangozeb_pam4,
  author={Aurangozeb and Dick, Carson R. and Mohammad, Maruf and Hossain, Masum},
  journal={IEEE Journal of Solid-State Circuits}, 
  title={{Sequence-Coded Multilevel Signaling for High-Speed Interface}}, 
  year={2020},
  volume={55},
  number={1},
  pages={27-37},
  doi={10.1109/JSSC.2019.2941016}}

@ARTICLE{pudfe,
  author={Kiran, Shiva and Cai, Shengchang and Luo, Ying and Hoyos, Sebastian and Palermo, Samuel},
  journal={IEEE Journal of Solid-State Circuits}, 
  title={{A 52-Gb/s ADC-Based PAM-4 Receiver With Comparator-Assisted 2-bit/Stage SAR ADC and Partially Unrolled DFE in 65-nm CMOS}}, 
  year={2019},
  volume={54},
  number={3},
  pages={659-671},
  doi={10.1109/JSSC.2018.2878850}}

\begin{IEEEbiography}[{\includegraphics[width=1in,height=1.25in,clip,keepaspectratio]{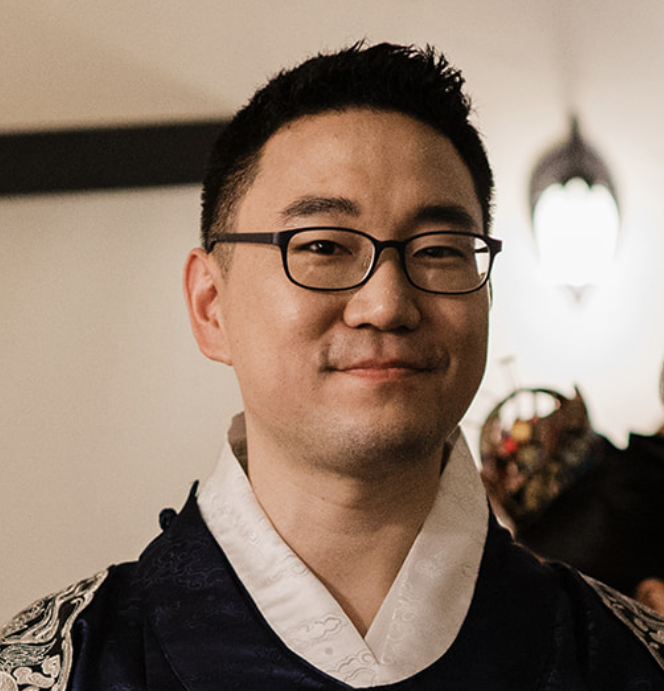}}]{Kunmo Kim} (Member, IEEE) received the B.S. degree from Texas A\&M University, College Station, TX, USA, in 2012, the M.S. degree from the California Institute of Technology, Pasadena, CA, USA, in 2014, and the Ph.D. degree in electrical engineering and computer sciences from the University of California, Berkeley, CA, USA, in 2025.

From 2014 to 2017, he was with Oracle, and from 2017 to 2018, he was with Apple, where he worked on SerDes circuit design. In 2024, he was with Marvell, Santa Clara, CA, USA, focusing on soft-decision forward-error-correction implementation and its impact on coding gain and latency. He is currently an R\&D Hardware Engineer with the Physical Layer Products Group, Broadcom, Irvine, CA, USA.

His research interests include nonlinear equalization for high-speed digital communication, statistical signal processing, and circuit design for high-speed serial links. He received the Qualcomm Innovation Fellowship in 2022.
\end{IEEEbiography}

\begin{IEEEbiography}[{\includegraphics[width=1in,height=1.25in,clip,keepaspectratio]{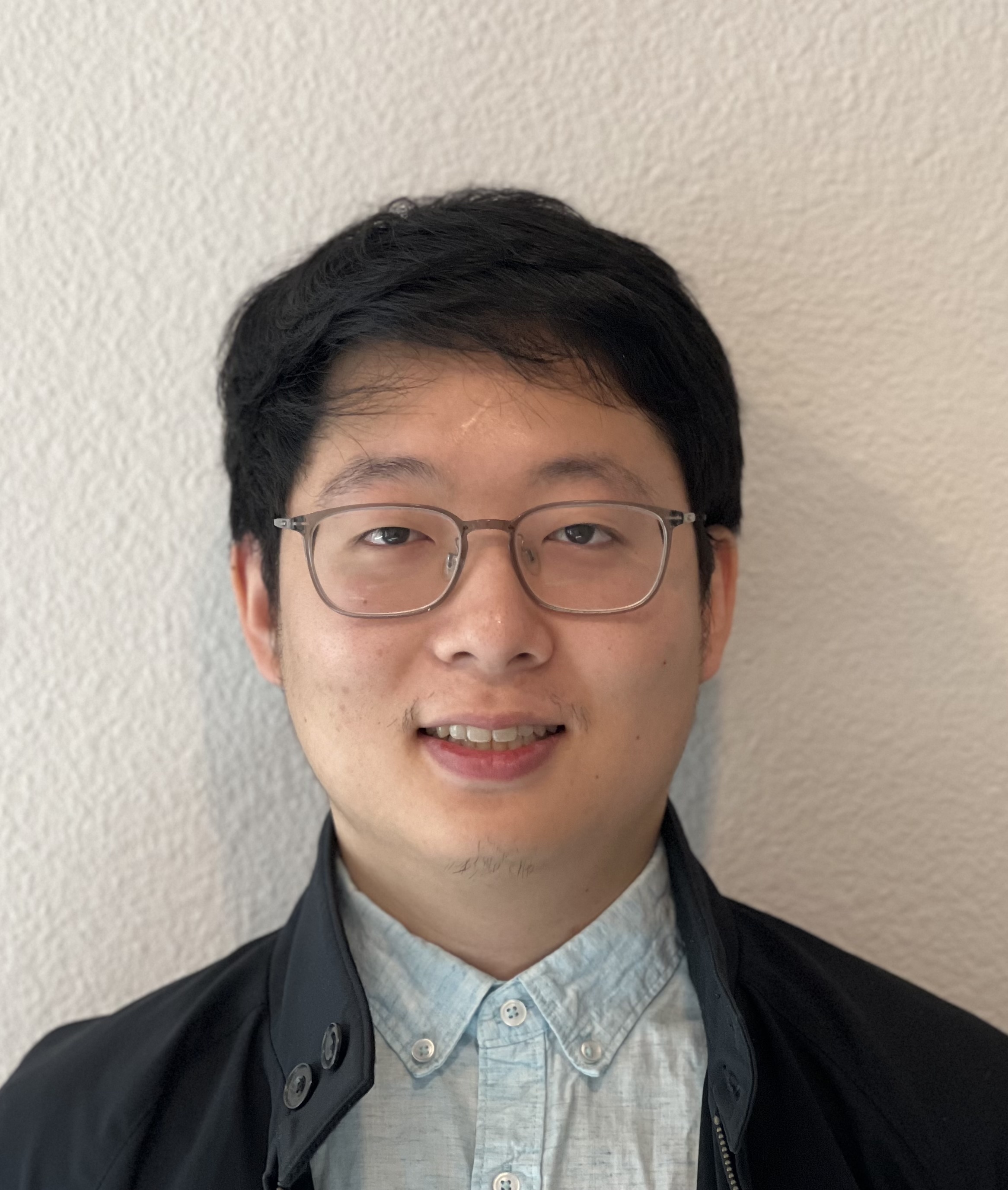}}]{Paul Kwon} (Member, IEEE) received the B.S. and Ph.D. degrees in electrical engineering and computer sciences from the University of California at Berkeley, Berkeley, CA, USA, in 2017 and 2023, respectively. He joined Tenstorrent (formerly Blue Cheetah Analog Design), Santa Clara, CA, USA, in 2023. He has previously interned with Apple. His research interests include die-to-die interconnects and the design automation of analog/mixed-signal circuits.
\end{IEEEbiography}

\begin{IEEEbiography}[{\includegraphics[width=1in,height=1.25in,clip,keepaspectratio]{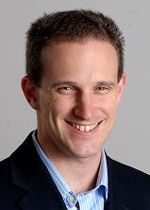}}]{Elad Alon} (Fellow, IEEE) received the B.S., M.S., and Ph.D. degrees in electrical engineering from Stanford University, Stanford, CA, USA, in 2001, 2002, and 2006, respectively.

He was a Co-Founder and the CEO of Blue Cheetah Analog Design, leading the company until its acquisition by Tenstorrent in June 2025. He is also an Adjunct Professor of electrical engineering and computer sciences with the University of California at Berkeley, Berkeley, CA, where he had previously served as a Full-Time Faculty Member and a Professor from 2007 to 2021. 

\end{IEEEbiography}

\begin{IEEEbiography}[{\includegraphics[width=1in,height=1.25in,clip,keepaspectratio]{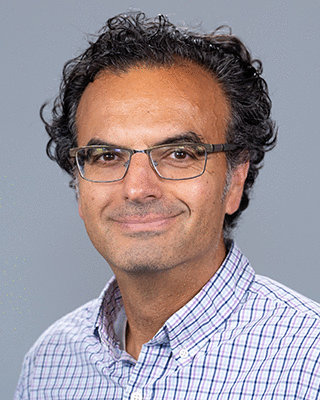}}]{Ali M. Niknejad} (Fellow, IEEE)  received the Ph.D. degree in electrical engineering from the University of California (UC), Berkeley, in 2000. He currently holds the Donald O. Pederson Distinguished Professorship chair in the EECS department at UC Berkeley, and he is a faculty co-director of the Berkeley Wireless Research Center (BWRC). He is also the Associate Director of the Center for Ubiquitous Connectivity (CUbiC) and served as the Associate Director for the Center for Converged TeraHertz Communications and Sensing (ComSenTer). He received the 2020 SIA/SRC University Research Award, recognized ``for noteworthy achievements that have advanced analog, RF, and mm-wave circuit design and modeling, which serve as the foundation of 5G+ technologies.'' 
%
\end{IEEEbiography}  

\end{document}